\documentclass[prb,twocolumn,superscriptaddress,nofootinbib,letterpaper]{revtex4-2}
\usepackage{amsmath,amssymb, amsfonts, amsthm}
\usepackage{bm}
\usepackage{tikz}
\usepackage{natbib}
\usepackage{booktabs}
\usepackage{pgffor}
\usepackage{mathrsfs} 
\usepackage{graphicx, color}
\usepackage{amstext, braket}
\usepackage{float}

\usetikzlibrary{decorations.pathmorphing} 

\usepackage[english]{babel}
\usetikzlibrary{calc}
\usetikzlibrary{positioning}
\usetikzlibrary{shapes.geometric}
\usetikzlibrary{arrows.meta}
\usetikzlibrary{positioning}
 
\usepackage{pythonhighlight}
\usepackage{setspace}

\usepackage[colorlinks=true,linkcolor=blue,citecolor=blue,urlcolor=blue]{hyperref}

\begin{document}
\title{Taxonomy of integrable and ground-state solvable models: Jastrow wave functions on graphs and parent Hamiltonians}

\author{Nilanjan Sasmal\href{https://orcid.org/0000-0002-9793-1439}{\includegraphics[scale=0.05]{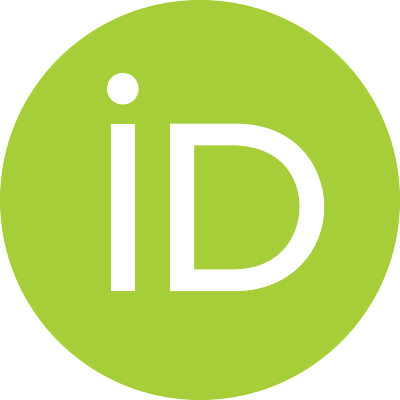}}}
\email{nilanjan.sasmal@uni.lu}
\address{Department of Physics and Materials Science, University of Luxembourg,
L-1511 Luxembourg, Luxembourg}
\author{Adolfo del Campo\href{https://orcid.org/0000-0003-2219-2851}{\includegraphics[scale=0.05]{orcidid.pdf}}}
\email{adolfo.delcampo@uni.lu}
\address{Department of Physics and Materials Science, University of Luxembourg,
L-1511 Luxembourg, Luxembourg}
\address{Donostia International Physics Center, E-20018 San Sebasti\'an, Spain}

\def\L{{\rm \hat{L}}}
\def\q{{\bf q}}
\def\l{\left}
\def\r{\right}
\def\te{\mbox{e}}
\def\d{{\rm d}}
\def\t{{\rm t}}
\def\K{{\rm K}}
\def\N{{\rm N}}
\def\H{{\rm H}}
\def\la{\langle}
\def\ra{\rangle}
\def\om{\omega}
\def\Om{\Omega}
\def\vep{\varepsilon}
\def\wh{\widehat}
\def\tr{{\rm Tr}}
\def\da{\dagger}
\def\iz{\left}
\def\zi{\right}
\newcommand{\beq}{\begin{equation}}
\newcommand{\eeq}{\end{equation}}
\newcommand{\beqa}{\begin{eqnarray}}
\newcommand{\eeqa}{\end{eqnarray}}
\newcommand{\intf}{\int_{-\infty}^\infty}
\newcommand{\into}{\int_0^\infty}

\begin{abstract}

We introduce a family of many-body systems of distinguishable continuous-variable particles in which interparticle interactions are set by the adjacency matrix of a graph. The ground-state wave function of such systems is of a generalized Jastrow form involving the product of pair-correlation functions over the edge set of the graph. These systems describe quantum fluids when the graph is complete, and the pair function has a well-defined permutation symmetry. In general, they provide the continuous-variable generalization of spin systems on graphs, with broken permutation symmetry. The corresponding parent Hamiltonian is shown to include (a) two-body interactions determined by the graph adjacency matrix and (b) three-body interactions over all possible 2-paths on the graph. Employing elements of graph theory, we chart the landscape of models, recovering known instances in the literature and providing numerous new examples of ground-state solvable models for which the system Hamiltonian, ground-state wave function, and corresponding energy eigenvalue are specified.

\end{abstract}

\maketitle

\section{Introduction}

Distinguishable degrees of freedom, such as spins, play a central role in the theory of quantum many-body systems. Since Bethe's solution of a one-dimensional interacting model \cite{Bethe31}, quantum spin Hamiltonians have served as paradigmatic settings for correlated matter and, more recently, for quantum computation and simulation. In quantum information science, distinguishable two-level systems (qubits) and their generalizations (qudits) provide a natural language for describing controllable many-body dynamics \cite{Nielsen00}. In parallel, networks of continuous-variable degrees of freedom, such as coupled oscillators, are widely used to model transport and entanglement properties in extended systems \cite{Dhar10, Eisert10}.

Bosons, fermions, and anyons are endowed with permutation symmetry, but particles are treated as distinguishable when corresponding to different species \cite{Calogero69, Moser75, Olshanetsky81, Olshanetsky83, Turbiner11}, as in the study of atomic mixtures \cite{Girardeau04, Harshman17}.  Freezing the spatial degree of freedom in systems of continuous variables provides a fruitful approach for deriving many-body spin systems, such as the celebrated Haldane-Shastry model \cite{Haldane88, Shastry88, Polychronakos93, Polychronakos06}. The breaking of permutation symmetry is also used in quantum Monte Carlo methods, e.g., when using Nosanow-Jastrow wave functions in the quantum theory of solids to effectively pin down particles at given coordinates \cite{Nosanow66}.  In such cases, permutation symmetry can then be restored by suitable symmetrization \cite{Cazorla07, Cazorla09, delcampo20}. Experimentally, particle ordering and addressability are increasingly routine in platforms such as ion chains, optical tweezer arrays, and Rydberg-based quantum simulators, where the relevant degrees of freedom are intrinsically labeled.

In this work, we consider networks of distinguishable single quantum systems with continuous variables. We focus on a class of models in which the ground-state wave function s set by the product of pair-correlation functions over a set of particle pairs. Such a structure in the ground-state wave function is known as the Jastrow form when it involves all possible particle pairs, i.e.,  those with permutation symmetry \cite{Sutherland04}. By contrast, in this work, we consider many-body quantum systems in which the ground-state wave function involves only pair-correlation terms at the edges of the graph. As it turns out, the parent Hamiltonian with such a ground-state involves pairwise interactions between the edges of the graph. In addition, it may involve three-body interactions spanning the set of 2-paths in the connectivity graph. The resulting Hamiltonians describe mixtures of distinguishable particles with continuous variables in a region of space, with interactions specified by the graph's adjacency matrix. These models can be further generalized to include a spatial distribution of particles on a lattice or graph by extending the Nosanow-Jastrow ansatz used in the quantum theory of solids to a graph-based formulation.

In what follows, we first derive the parent Hamiltonian associated with a Graph-Jastrow wave function (GJW) in arbitrary spatial dimensions and subsequently specialize to one-dimensional systems, where the structure simplifies considerably, and a rich taxonomy emerges. We begin with the case of all-to-all pairwise interactions corresponding to a complete graph. In this limit, we recover a broad and important class of exactly solvable models \cite{delcampo20, Beau21}, including the Calogero-Sutherland family \cite{Calogero71, Sutherland71}, trigonometric variants \cite{Sutherland04, KuramotoKato09}, and the Lieb-Liniger gas \cite{LL63, L63, McGuire64, KBI97, Takahashi99}, among others. When the pair-correlation function possesses a definite parity, the resulting ground state is fully (anti)symmetric under particle exchange, thereby restoring permutation symmetry and describing bosonic or fermionic quantum fluids.

We then extend the construction to arbitrary connectivity graphs and use elementary tools from graph theory to organize the resulting models. When the graph is a path or cycle, the formalism yields a natural generalization of the Jain-Khare model \cite{JainKhare99, Auberson01, Basu-Mallick01, Ezung05, Enciso05, Jain06} for arbitrary pair functions, featuring inverse-square nearest-neighbor interactions supplemented by an induced attractive three-body term. For regular graphs with finite coordination numbers, one recovers and generalizes truncated Calogero-Sutherland-type models \cite{Pittman17, Tummuru17, Baradaran18, Yadav19}.

Beyond unifying these known examples, our main objective is taxonomic. For each graph family and each admissible pair function $f$, the framework explicitly specifies the parent Hamiltonian, the exact ground-state wave function, and the corresponding ground-state energy. Moreover, we show how standard graph operations such as joins and graph products systematically generate new composite models with transparent physical interpretations, including continuum analogs of ``system + environment'' constructions familiar from impurity and decoherence problems.

\section{Quantum Many-body wave functions on Graphs}
We are interested in describing quantum many-body systems of $N$ particles, each of which is associated with a vertex of the graph. The interaction between any two particles is associated with an edge, so the resulting graph describes the interactions between particles. This is a common approach, e.g., in the study of networks of oscillators. Our interest is in the identification of more general quasi-solvable many-body quantum models for which several exact properties can be derived, such as the ground-state wave function, its energy, and the parent Hamiltonian.

To this end, we introduce a finite set $V=\{1, 2,\dots, N\}$ and consider a classical graph $\mathcal{G}(V, E)$ with $N$ vertices and $M$ edges, where $V$ and $E$ denote the sets of vertices and edges of $\mathcal{G}$, respectively. The set of edges is composed of pairs $(i,j)$ labeling an edge that starts at vertex $i$ and ends at vertex $j$. Such a graph can be described by the adjacency matrix  $A=A(\mathcal{G})$ with elements $A_{ij}=1$ if $(i, j)\in E$ and  $A_{ij}=0$ otherwise. 

In the description of quantum many-body networks, we endow each vertex with a quantum-mechanical particle in space dimension $D$, i.e., with coordinate $\vec{r}_i\in\mathbb{R}^D$.  We shall focus on ground-state wave functions with a Jastrow form and associated pair correlation functions $f_{ij}=f(\|\vec{r}_i-\vec{r}_j\|)$ of particles $i$ and $j$ with an edge $(i,j)= (j, i)$. The graph accounting for pair correlations in the ground-state wave function is thus undirected, and its adjacency matrix satisfies $A_{ii}=0$, $A_{ij}=A_{ji}$. As we shall see, there is a one-to-one correspondence between the graph of pair correlations in the generalized Jastrow ansatz and the graph accounting for two-body interactions in the corresponding parent Hamiltonian. The presence of pair correlation functions in the ground-state wave function and of two-body interactions in the system Hamiltonian is determined by the graph's adjacency matrix.

To make use of the adjacency matrix, we shall thus avoid the double counting and define the set $E$ as that of all edges $(i,j)$ of $\mathcal{G}$ with the edge $i<j$. We shall exclude self-interaction between particles, which is tantamount to considering an antireflexive graph with no $(i, i)$ edges. The cardinalities of $V$ and $E$ are denoted by $|V|$ and $|E|$, respectively. Given that we consider one particle in each node of the graph, $|V|$ equals the total particle number $N$ and sets the order of the graph. The number of edges $|E|=M$ is also known as the size of the graph. 

The Hilbert space of a many-body system on a graph $\mathcal{G}(V, E)$ is spanned by the set of square-integrable functions
\beqa
\mathcal{H}_{\mathcal{G}}=L^2(\mathbb{R}^{DN},d\mathbf{r})=\bigotimes_{i\in V}L^2(\mathbb{R}^D,d^Dr_i),
\eeqa
with $\mathbf{r}=(\vec{r}_i)_{i\in V}$ and $\vec{r}_i = (x_i^1, x_i^2, \cdots, x_i^D) \in \mathbb{R}^D$ denotes the coordinate of the $i$th particle, i.e., that on the node $i$. However, as in the study of conventional Jastrow wave functions, it is often convenient to focus on wave functions containing only the pairwise product of pair functions, which are not square integrable. The inclusion of one-body functions in the Jastrow ansatz can be justified in the presence of an external potential, and exact identities relating quantum states with and without confinement are known \cite{delcampo20, Beau20, Beau21, Yang22}. Similar identities also apply to the corresponding Hamiltonians and their eigenvalues.

We focus exclusively on interacting systems associated with a connected graph such that a path (correlations) exists between any two vertices (particles). In doing so, we exclude mixtures of non-interacting species. However, the latter can be accounted for by a straightforward generalization of our construction to disconnected graphs.

\section{Parent Hamiltonian of Graph Jastrow wave functions}

In this section we derive the parent Hamiltonian of a Graph-Jastrow wave function in arbitrary spatial dimension $D$, showing that it splits into a kinetic part, a two-body interaction supported on the edges of the graph, and a three-body interaction supported on its 2-paths.

Specifically, we are interested in finding the parent Hamiltonian $\hat{H}_0$ with an $N$-particle ground-state wave function of the Graph-Jastrow form,
\beqa
\label{Psi0free}
\Phi_0(\vec{r}_1,\dots,\vec{r}_N) = \prod_{(i,j)\in E}f(r_{ij}) = \prod_{i<j}e^{A_{ij} \ln f(r_{ij})},
\eeqa
where $ r_i = ||\vec{r}_i|| = \sqrt{\sum_{\mu=1}^D (x_i^{\mu})^2} $ is the Euclidean norm, $A_{ij}$ are the elements of the adjacency matrix $A$, and the product runs over all edges in the graph (with $i<j$) and  $r_{ij}=\|\vec{r}_i-\vec{r}_j\|$. Thus, we consider the pair function $f(r_{ij})$ to depend exclusively on the relative distance between particles.

In $D$ spatial dimensions, we can write the single-particle Laplacian in hyperspherical coordinates as 
\begin{equation}\label{p1.61}
    \begin{split}
        \nabla^2 & =  \frac{1}{r^{D-1}}\frac{\partial}{\partial r}\bigg(r^{D-1}\frac{\partial}{\partial r}\bigg) + \frac{\Delta^{S^{D-1}}(\theta_1, \theta_2, \cdot\cdot\cdot, \varphi)}{r^2}  ,
    \end{split}
\end{equation}
to determine the parent Hamiltonian $\hat{H}_0$, satisfying $\hat{H}_0\Phi_0=0$, we first note that the kinetic  energy operator in $D$-spatial dimensions can be written in hyperspherical coordinates as
\beqa
\label{Tddim}
\hat{T}&=&-\frac{\hbar^2}{2m}\sum_{i=1}^N\left[\frac{1}{r_i^{D-1}}\frac{\partial}{\partial r_i}\bigg(r_i^{D-1}\frac{\partial}{\partial r_i}\bigg)+\frac{\Delta^{S^{D-1}}}{r_i^{2}}\right], \ \ 
\eeqa
where the Laplace-Beltrami operator on the sphere $S^{D-1}$ is denoted by $\Delta_i^{S^{D-1}}$. Noting that $\Delta_i^{S^{D-1}}\Phi_0=0$, we conclude that the parent Hamiltonian of the GJW involves exclusively two-body and three-body interactions and takes the form
\beqa
\label{Hgraph}
\hat{H}_0=-\frac{\hbar^2}{2m}\sum_{i = 1}^N \Delta_i+ \hat{V}_2+\hat{V}_3,
\eeqa
where the two- and three-body potentials are given by
\begin{eqnarray}
\hat{V}_2&=&\frac{\hbar^2}{m}\sum_{i<j} A_{ij}\bigg[  \frac{f''_{ij}}{f_{ij}} + \frac{(D-1)}{r_{ij}} \frac{f'_{ij}}{f_{ij}}\bigg],\\
\hat{V}_3&=&\frac{\hbar^2}{2m}\sum_{i=1}^N \sum_{k\neq j \neq i}A_{ij}A_{ik} (\hat{r}_{ij}\cdot\hat{r}_{ik} ) \frac{f'_{ij}}{f_{ij}} \frac{f'_{ik}}{f_{ik}}, 
\end{eqnarray}
where $\hat{r}_{ij}:=(\vec{r}_i-\vec{r}_j)/r_{ij}$ and the sum subindex $k\neq j \neq i$  indicates the sum over the three indices $i,j,k$ excluding any repetition. 
Here, $f'$ and $f''$ denote the first and second spatial derivatives of $f(r_{ij})=f_{ij}$, respectively. 

The unit vector $\hat{r}_{ij}$ is not defined on the coincidence hyperplanes $r_{ij}=0$, but this does not pose a difficulty, when either the pair function vanishes at $r_{ij}=0$, so that $\Phi_0$ acquires a node on the coincidence set, or the two-body potential $\hat{V}_2$ develops a repulsive $1/r_{ij}^2$-type barrier that renders the coincidence set dynamically inaccessible \cite{Beau21}. 
In the subsequent sections, we focus on systems in one spatial dimension, in which no difficulty arises because the role of the unit vector is replaced by the sign of the relative coordinate, defined to be zero at coincidence \cite{delcampo20}.

In $\hat{V}_3$, the index $i$ plays the role of the central vertex of the 2-path $j$--$i$--$k$; since the graph is undirected ($A_{ij}=A_{ji}$) and the kernel $A_{ij}A_{ik}(\hat{r}_{ij}\!\cdot\!\hat{r}_{ik})(f'_{ij}/f_{ij})(f'_{ik}/f_{ik})$ is symmetric under $j\leftrightarrow k$, each unordered 2-path is counted twice and the $1/2$ prefactor compensates this double counting; the result is therefore independent of any labeling of the vertices.
\begin{table*}[htbp]
\setlength{\tabcolsep}{4pt}
\begin{tabular}{@{}llll@{}} 
\scriptsize Pair function 
& \scriptsize Two-body potential: $\hat{V}_2$ 
& \scriptsize Three-body potential: $\hat{V}_3$ 
& \scriptsize Model/Interactions \\ 

\scriptsize $f(x_{ij})$
& \scriptsize $\displaystyle \frac{\hbar^2}{2m} \sum_i \sum_{j\neq i} A_{ij}\frac{f''_{ij}}{f_{ij}}$
& \scriptsize $\displaystyle \frac{\hbar^2}{2m} \sum_i \sum_{k\neq j\neq i} 
A_{ij}A_{ik}\frac{f'_{ij}}{f_{ij}}\frac{f'_{ik}}{f_{ik}}$
& \scriptsize Arbitrary interaction \\ 

\scriptsize $|x_{ij}|^g$
& \scriptsize $\displaystyle \frac{\hbar^2 g(g-1)}{m} \sum_{i<j} \frac{A_{ij}}{|x_{ij}|^2}$
& \scriptsize $\displaystyle \frac{\hbar^2 g^2}{m} \sum_{i<j<k}
\!\left(\frac{A_{ij}A_{ik}}{x_{ij}x_{ik}}
-\frac{A_{ji}A_{jk}}{x_{ij}x_{jk}}
+\frac{A_{ki}A_{kj}}{x_{ik}x_{jk}}\right)$
& \scriptsize Inverse square \\ 

\scriptsize $\exp(g|x_{ij}|)$
& \scriptsize $\displaystyle \frac{\hbar^2 g}{m} \sum_{i<j} A_{ij}\,[g+2\delta(x_{ij})]$
& \scriptsize $\displaystyle \frac{\hbar^2 g^2}{m} \sum_{i<j<k}
(\tilde A_{ij}\tilde A_{ik}+\tilde A_{ji}\tilde A_{jk}+\tilde A_{ki}\tilde A_{kj})$
& \scriptsize Contact \\ 

\scriptsize $\exp(g|x_{ij}|^2)$
& \scriptsize $\displaystyle \frac{2\hbar^2 g}{m} \sum_{i<j} A_{ij}(1+2g|x_{ij}|^2)$
& \scriptsize $\displaystyle \frac{2\hbar^2 g^2}{m} \sum_{i<j<k}
(A_{x_{ij}}A_{x_{ik}}+A_{x_{ji}}A_{x_{jk}}+A_{x_{ki}}A_{x_{kj}})$
& \scriptsize Coupled oscillators \\ 

\scriptsize $|\sinh(x_{ij}/\ell)|^g$
& \scriptsize $\displaystyle \frac{\hbar^2 g}{m\ell^2} \sum_{i<j} A_{ij}
\!\left[g+\frac{(g-1)}{\sinh^2(x_{ij}/\ell)}\right]$
& \scriptsize $\displaystyle \frac{\hbar^2 g^2}{2m\ell^2} \sum_{i<j<k}
(A_{ct_{ij}}A_{ct_{ik}}+A_{ct_{ji}}A_{ct_{jk}}+A_{ct_{ki}}A_{ct_{kj}})$
& \scriptsize Hyperbolic \\ 

\end{tabular}
\caption{Quantum many-body systems associated with an arbitrary graph $G_N$ and the ground-state GJW $\Phi_0$. The examples are chosen such that the three-body potential $\hat{V}_3$ reduces to a constant or a two-body contribution on a complete graph. Definitions: $\tilde A_{pq}=A_{pq}\mathrm{sgn}(x_{pq})$, $A_{x_{pq}}=A_{pq}x_{pq}$, and $A_{ct_{pq}}=A_{pq}\coth(x_{pq}/\ell)$.}
\label{GN_models}
\end{table*}

Note that by construction $\hat{H}_0\Phi_0=0$. As we next show, this Hamiltonian is positive semidefinite, $\hat{H}_0\geq 0$, whence it follows that $\Phi_0$ is the true ground state.  
To this end, let us define
\begin{eqnarray}
Q_i&:=&\frac{\hbar}{\sqrt{2m}}\Big(\nabla_i-\sum_{j\neq i} A_{ij}\frac{f'_{ij}}{f_{ij}}\hat{r}_{ij}\Big),\\
Q_i^\dag&:=&\frac{\hbar}{\sqrt{2m}}\Big(-\nabla_i-\sum_{j\neq i} A_{ij}\frac{f'_{ij}}{f_{ij}}\hat{r}_{ij}\Big),
\end{eqnarray}
with $Q_i\Phi_0=0$ for all $i$.
Then the parent Hamiltonian admits the exact factorization
\begin{equation}
\hat H_0=\sum_{i=1}^N Q_i^\dag Q_i.
\end{equation}
In particular, for any state $\Phi$
for which the integration by parts that converts $Q_i^\dagger$ into the formal adjoint of $Q_i$ is justified, one has 
\begin{equation}
\langle\Phi,\hat H_0\Phi\rangle=\sum_{i=1}^N \|Q_i\Phi\|^2\ge 0,
\end{equation}
so $\hat H_0$ is positive semidefinite, and $\Phi_0$ is a zero-energy ground state. In many specific instances, however, $\hat{V}_2$ and $\hat{V}_3$ include constant terms that can be absorbed in a nonzero ground-state energy $E_0$, that is, by setting $(\hat{H}_0-E_0)\Phi_0=0$.

We further note that,  as $Q_i\Phi_0=0$, there is an infinite family of Hamiltonians with $\Phi_0$ as ground state, including those of the form
\begin{eqnarray}
    \hat{H}_\kappa & = & \sum_{i,j} \kappa_{ij} Q_i^\dagger Q_j, \quad \kappa_{ij} = \kappa_{ji}^* \in \mathbb{C}.
\end{eqnarray} 
Indeed, any such $\hat{H}_\kappa$ is still a positive semidefinite Hamiltonian with $\Phi_0$ as ground state.

% \vspace{-0.2 cm}

\section{Parent Hamiltonian of a Graph-Jastrow wave function Under Confinement}
One can generalize the above construction by considering the possibility that each particle experiences a one-body confinement potential $\textsl{g}(\vec{r}_i)$. We can write the Ansatz for the ground-state of the many-body Hamiltonian $\hat{H}$ as 
\beqa
\label{Psi0pot}
\Psi_0(\vec{r}_1, \cdot\cdot\cdot,\vec{r}_N) = \prod_{i \in V} \textsl{g}(\vec{r}_i) \prod_{i<j}e^{A_{ij} \ln f(r_{ij})}.
\eeqa

Following \cite{delcampo20,Beau21}, by  explicit evaluation of the action of the Laplacian on $\Psi_0$, one finds the parent Hamiltonian, satisfying $\hat{H} \Psi_0 = 0$, to be given by

\begin{align}
    \hat{H} & = \hat{H}_0+  \hat{V}_{1} + \hat{V}_{\rm 2LL}, \\
    \label{EqV1}
    \hat{V}_{ 1} &= \frac{\hbar^2}{2m} \sum_{i=1}^N \l[\frac{(D-1)}{r_i}\frac{\textsl{g}'_i}{\textsl{g}_i}+\frac{\textsl{g}''_i}{\textsl{g}_i}\r], \\ 
    \label{EqV2LL}
\hat{V}_{\rm 2LL} & = 
\frac{\hbar^2}{2m} \sum_{i<j} A_{ij} \frac{f'_{ij}}{f_{ij}} \hat{r}_{ij} \cdot \bigg[  \frac{\textsl{g}'_i}{\textsl{g}_i}  \hat{r}_i  - \frac{\textsl{g}'_j}{\textsl{g}_j}  \hat{r}_j \bigg].
\end{align}

Thus, the inclusion of the one-body function $\textsl{g}(\vec{r}_i)=\textsl{g}_i$ in the GJW gives rise to an external one-body potential $\hat{V}_{\rm 1} $ as well as pairwise interactions $\hat{V}_{\rm 2LL}$ that are generally long range. As an example, choosing a Gaussian wave function 
\beqa
\textsl{g}(\vec{r}_i)=\exp\left(-\frac{m\omega}{2\hbar}r_i^2\right),
\eeqa
gives rise to a harmonic trap as a confining potential 
\beqa
\hat{V}_{1}=\sum_{i=1}\frac{1}{2}m\omega^2 r_i^2 - \frac{D}{2}N\hbar\omega,
\eeqa
and a long-range potential
\beqa
\hat{V}_{\rm 2LL} =-\frac{\hbar\omega}{2} \sum_{i<j} A_{ij} \frac{f'_{ij}}{f_{ij}} {r}_{ij}.
\eeqa
These results have a direct correspondence in models involving all-to-all pairwise interactions \cite{Beau21}.

\section{Parent Hamiltonian of Graph Jastrow wave functions in One Dimension}
We shall mostly focus on the one-dimensional (1D) case with $D=1$, where an important simplification occurs. We are interested in finding the parent Hamiltonian $\hat{H}_0$ with a ground-state of the GJW form 
\beqa
\label{Psi0free}
\Phi_0(x_1,\cdots,x_N) = \prod_{(i,j)\in E} f_{ij} = \prod_{i<j} e^{A_{ij} \ln f_{ij}},
\eeqa
where $x_{ij}=x_{i,j}=x_i-x_j$, $A_{ij}=A_{ji}$, and $f_{ij}=f_{ji}$.  The parent Hamiltonian $\hat{H}_0$ satisfying $\hat{H}_0\Phi_0=0$ simplifies to

\begin{equation}\label{V3Eq}
    \begin{split}
     \hat{H}_0 & = -\frac{\hbar^2}{2m} \sum_{i=1}^N \frac{\partial^2}{\partial x_i^2} + \hat{V}_2 + \hat{V}_3,  \\
        \hat{V}_2 & =   \frac{\hbar^2}{2m} \sum_{i=1}^N \sum_{j\neq i} A_{ij} \frac{f''_{ij}}{f_{ij}}, \\ \hat{V}_3 & = \frac{\hbar^2}{2m}\sum_{i=1}^N \sum_{k\neq j \neq i} A_{ij}A_{ik}  \frac{f'_{ij}}{f_{ij}} \frac{f'_{ik}}{f_{ik}}. 
    \end{split}
\end{equation}

In the standard Bijl-Jastrow form, in which the product in Eq. (\ref{Psi0free}) extends over all possible pairs of particles, the quantum exchange statistics is encoded in the pair correlation function. When the pair correlation function is symmetric (antisymmetric), $\Phi_0$ describes bosons  (fermions). One-dimensional anyons can be accommodated using anyon-fermion and anyon-boson dualities \cite{Kundu99, Girardeau06, Batchelor06, delcampo08, mackel2022quantum}. In the following, we shall be interested in graphs with restricted interactions and assume that particles are distinguishable. Exchange statistics may then be recovered by imposing symmetrization ``by hand", as in the description of quantum solids using Nosanow-Jastrow ans\"atze \cite{Nosanow66, Cazorla07, Cazorla09, delcampo20}.

Equations (\ref{Psi0free}) and (\ref{V3Eq}) depend explicitly on the adjacency matrix of the graph that determines the ground-state pair correlations, as well as the pairwise interactions between particles.  In what follows, we shall classify instances of this family of systems according to the adjacency matrix; see Table \ref{GN_models}.

\begin{table*}[htbp]
\setlength{\tabcolsep}{4pt}
\begin{tabular}{@{}llll@{}} 
\scriptsize{Pair function} & \scriptsize{Two-body potential: $\hat{V}_2$} & \scriptsize{Three-body potential: $\hat{V}_3$} & \scriptsize{Model/Interactions} \\  \scriptsize{$f(x_{ij})$}  & \scriptsize{$\displaystyle \frac{\hbar^2}{m}  \sum_{ i < j} \frac{f''_{ij}}{f_{ij}}$} &  \scriptsize{$\displaystyle \frac{\hbar^2}{m}\sum_{i<j<k}\left[\frac{f'_{ij}f'_{ik}}{f_{ij}f_{ik}}
-\frac{f'_{ij}f'_{jk}}{f_{ij}f_{jk}}+\frac{f'_{ik}f'_{jk}}{f_{ik}f_{jk}}\right] $} & \scriptsize{Arbitrary interaction \cite{delcampo20}} \\  
\scriptsize{$\displaystyle |x_{ij}|^g$ } &  \scriptsize{$\displaystyle \frac{\hbar^2}{m}\sum_{i<j}\frac{g(g-1)}{|x_{ij}|^2}$} & \scriptsize{$0$} & \scriptsize{Calogero-Moser \cite{Calogero69,Calogero71,Moser75}} \\  
\scriptsize{$\displaystyle \exp(g|x_{ij}|)$} &  \scriptsize{$\displaystyle \frac{\hbar^2g^2}{2m}{N(N-1)} + \frac{2\hbar^2g}{m}\sum_{i<j}\delta(x_{ij})$} & \scriptsize{$\displaystyle\frac{\hbar^2g^2}{6m} N(N-1)(N-2)$} & \scriptsize{Lieb-Liniger  \cite{LL63, L63, McGuire64, Yang1967, Li:1995wr}}\\
 \scriptsize{$\displaystyle \exp(g|x_{ij}|^2)$ }                 &  \scriptsize{$\displaystyle\frac{\hbar^2g}{m}N(N-1) +\frac{4\hbar^2g^2}{m} \sum_{i<j}|x_{ij}|^2$} & \scriptsize{$\displaystyle \frac{2\hbar^2g^2}{m}(N-2)\sum_{i<j}|x_{ij}|^2$} &  \scriptsize{Coupled  oscillators} \\
 \scriptsize{$\displaystyle |\sinh(x_{ij}/\ell)|^g$}                   &  \scriptsize{$\displaystyle \frac{\hbar^2 g^2 }{2m\ell^2}N(N-1)+\frac{\hbar^2}{m\ell^2}\sum_{i<j}\frac{g(g-1)}{{\rm sinh}^2(x_{ij}/\ell)}$} & \scriptsize{$\displaystyle \frac{\hbar^2g^2}{6m\ell^2}N(N-1)(N-2)$} & \scriptsize{Hyperbolic}   \\ 
\end{tabular}
\caption{Quantum many-body systems associated with a complete graph $K_N$ and ground-state wave function $\Phi_0=\prod_{i<j}f(x_{ij})$. The examples considered are such that the three-body potential $\hat{V}_3$ reduces to a constant or to a two-body contribution \cite{delcampo20}.}
\label{KNmodels}
\end{table*}

\section{$K_N$ complete graph models: All-to-all pairwise interactions }

We first specialize the construction to the complete graph $K_N$, in which every pair of particles is connected, i.e., they interact with each other. This corresponds to the standard Bijl--Jastrow setting, in which we recover well-known integrable many-body models as particular cases.

Standard quantum fluids have interactions between all possible neighbors, and the ground-state GJW takes the Bijl-Jastrow form
\beqa
\label{BJGS}
\Phi_0(x_1,\cdots,x_N)=\prod_{i<j}f_{ij},
\eeqa
which can be associated with the adjacency matrix
\beqa
\label{Achain}
A_{ij}=1-\delta_{ij},
\eeqa
the latter corresponds to a complete graph $K_N$ where the number of edges is  $|E|=\binom{N}{2}=N(N-1)/2$. 

The complete family of models with a wave function of Bijl-Jastrow form has been discussed, e.g., in \cite{delcampo20}. The parent Hamiltonian (\ref{Hgraph}) with ground state (\ref{BJGS}) involves   a two-body potential with all-to-all pairwise interactions,
\beqa
\label{V2KN}
\hat{V}_2=\frac{\hbar^2}{m}  \sum_{ i < j} \frac{f''_{ij}}{f_{ij}}.
\eeqa

The three-body interaction spans over all possible triplets, 
\beqa
\label{V3KN}
\hat{V}_3=\frac{\hbar^2}{m}\sum_{i<j<k}\left[\frac{f'_{ij}f'_{ik}}{f_{ij}f_{ik}}
-\frac{f'_{ij}f'_{jk}}{f_{ij}f_{jk}}+\frac{f'_{ik}f'_{jk}}{f_{ik}f_{jk}}\right],
\eeqa
and the number of paths of path length 2 is $p_2(K_N)=N(N-1)(N-2)/2$. Here and throughout, $p_2(\mathcal{G})$ denotes the number of length-2 paths in the graph $\mathcal{G}$, i.e., the number of ordered triples $(i,j,k)$ with $i\neq k$ such that $(i,j),(j,k)\in E$; equivalently, $p_2(\mathcal{G})=\sum_{v\in V}d(v)\,(d(v)-1)$, where $d(v)$ is the degree of vertex $v$ \cite{Godsil01}. This is the natural counting object for the three-body terms in $\hat{V}_3$.
Some relevant instances in this category are shown in Table \ref{KNmodels}. The ground-state energy becomes $E_0=-\frac{\hbar^2g^2}{6m}N(N^2-1)$, which is exactly the McGuire soliton energy in free space \cite{McGuire64}, as a result of the constant contributions from $\hat{V}_2$ and $\hat{V}_3$, when we choose our pair function to be $\exp(g|x_{ij}|)$.

\vspace{-0.5 cm}

\section{Path $P_N$ and Cycle $C_N$ Graph Nodes with Nearest Neighbor Interactions}
We next consider a ground-state with a GJW form restricted to nearest neighbors with an arbitrary pair function $f_{i,i+1} = f(x_{i,i+1})$, 
\begin{equation}\label{k-NN1}
    \Phi_0(x_1,\cdots, x_N)    = \prod_{i=1}^N f_{i,i+1} .
    \end{equation}
    
Systems of $N$ particles with nearest-neighbor interactions can be associated with path and cycle graphs, $P_N$ and $C_N$, respectively. We recall that the path graph $P_N$ is the connected graph on $N$ vertices with edge set $E=\{(i,i+1):i=1,\dots,N-1\}$, i.e., a simple linear chain, while the cycle graph $C_N$ is obtained from $P_N$ by adding the single edge $(N,1)$ \cite{Godsil01}, corresponding to periodic boundary conditions. They satisfy
\begin{equation}
    \left.
    \begin{aligned}
        |E| & = N-1,\quad  p_2(G) = N-2, \quad  {\rm for \ } P_N \\ 
         |E| & = N,\quad  p_2(G) = N, \quad  {\rm for \ } C_N
    \end{aligned}
    \right\}. 
\end{equation}

By construction, the wave function (\ref{k-NN1}) breaks the permutation symmetry and assumes the distinguishability of particles, so that they can be labeled (ordered) and that pair correlations are restricted to nearest neighbors. Symmetry can be restored by explicit symmetrization. Correlations in the wave function \eqref{k-NN1} can be associated with the adjacency matrix
\begin{equation}
\left.
\begin{aligned}
A_{ij} & = \delta_{i, j+1} + \delta_{i,j-1}, \quad  {\rm for \ } P_N \\
A_{ij} & = \delta_{i, (j+1){\rm \  mod \ }N} + \delta_{i,(j-1){\rm \  mod \ }N}, \quad  {\rm for \ } C_N
\end{aligned}
\right\}.
\end{equation}

\begin{table*}[htbp] 
\setlength{\tabcolsep}{4pt}
\begin{tabular}{@{}llll@{}} 
\scriptsize{Pair function} & \scriptsize{Two-body potential: $\hat{V}_2^{(P_N)}$} & \scriptsize{Three-body potential: $\hat{V}_3^{(P_N)}$} & \scriptsize{Model/interactions} \\ 
\scriptsize{$f(x_{i,i+1})$} 
& \scriptsize{$\displaystyle \frac{\hbar^2}{m} \sum_{i=1}^{N-1} 
    \frac{f''_{i,i+1 }}{f_{i,i+1}}   $}
& \scriptsize{{$\displaystyle -\frac{\hbar^2}{m} \sum_{i=2}^{N-1} \frac{f'_{i-1,i}}{f_{i-1,i}}
    \frac{f'_{i,i+1}}{f_{i,i+1}}$} } 
& \scriptsize{Arbitrary interaction} \\ 

\scriptsize{$|x_{i,i+1}|^g$} & \scriptsize{$\displaystyle \frac{\hbar^2}{m}  \sum_{i=1}^{N-1}  \frac{g(g-1)}{|x_{i,i+1}|^2} $} & \scriptsize{$\displaystyle - \frac{\hbar^2}{m} \sum_{i=2}^{N-1} \frac{g^2}{x_{i-1,i} \ x_{i,i+1}} $} & \scriptsize{Jain-Khare \cite{JainKhare99}} \\ 

\scriptsize{$\exp(g|x_{i,i+1}|)$} & \scriptsize{$\displaystyle \frac{\hbar^2g}{m} \bigg[  g (N-1) +  2 \sum_{i=1}^{N} \delta(x_{i,i+1}) \bigg]  $} & \scriptsize{ $ \displaystyle \frac{\hbar^2g^2}{m} (N-2) $ } & \scriptsize{Contact \cite{Yang1967,Li:1995wr}}  \\ 

\scriptsize{$\exp(g|x_{i,i+1}|^{2})$} & \scriptsize{$ \displaystyle \frac{2\hbar^2 g}{m} \bigg[  (N-1) + 2g \sum_{i=1}^{N} |x_{i,i+1}|^{2} \bigg] $} & \scriptsize{$ \displaystyle - \frac{4\hbar^2g^2}{m} \sum_{i=2}^{N-1} x_{i-1,i} \ x_{i,i+1}$} & \scriptsize{Coupled oscillators} \\ 

\scriptsize{$|\sinh{(x_{i,i+1}/\ell)}|^g$} & \scriptsize{$\displaystyle \frac{\hbar^2g}{m\ell^2} \bigg[ (N-1) + (g-1) \sum_{i=1}^{N-1} \coth^2{(x_{i,i+1}/\ell)} \bigg] $} & \scriptsize{$ \displaystyle - \frac{\hbar^2g^2}{m\ell^2} \sum_{i=2}^{N-1} \coth{(x_{i-1,i}/\ell)} \coth{(x_{i,i+1}/\ell) } $} & \scriptsize{Hyperbolic} \\ 

\end{tabular}
\caption{Quantum many-body systems associated with a path graph $P_N$ and ground-state wave function $\Phi_0=\prod_{i=1}^{N-1}f(x_{i,i+1})$. The three-body contribution becomes non-trivial even in the cases where it reduces to a constant or a two-body term when the graph is complete.}
\label{PNmodels}
\end{table*}
\begin{table*}[htbp] 
\setlength{\tabcolsep}{4pt}
\begin{tabular}{@{}llll@{}} 
\scriptsize{Pair function} & \scriptsize{Two-body potential: $\hat{V}_2^{(C_N)}$} & \scriptsize{Three-body potential: $\hat{V}_3^{(C_N)}$} & \scriptsize{Model/interactions} \\ 
\scriptsize{$f(x_{i,i+1})$} 
& \scriptsize{$\displaystyle  \frac{\hbar^2}{m} \sum_{i=1}^{N} 
    \frac{f''_{i,i+1 }}{f_{i,i+1}}    $}
& \scriptsize{ $\displaystyle -\frac{\hbar^2}{m} \sum_{i=1}^{N} \frac{f'_{i-1,i}}{f_{i-1,i}}
    \frac{f'_{i,i+1}}{f_{i,i+1}}  $} 
& \scriptsize{Arbitrary interaction } \\ 

\scriptsize{$|x_{i,i+1}|^g$} & \scriptsize{$\displaystyle \frac{\hbar^2}{m}  \sum_{i=1}^{N}  \frac{g(g-1)}{|x_{i,i+1}|^2} $} & \scriptsize{$\displaystyle -\frac{\hbar^2}{m}  \sum_{i=1}^{N} \frac{g^2}{x_{i-1,i} \ x_{i,i+1}} $} & \scriptsize{Jain-Khare \cite{JainKhare99}} \\ 

\scriptsize{$\exp(g|x_{i,i+1}|)$} & \scriptsize{$\displaystyle \frac{\hbar^2g}{m} \bigg[  g N +  2 \sum_{i=1}^{N} \delta(x_{i,i+1}) \bigg]  $} & \scriptsize{ $ \displaystyle \frac{\hbar^2g^2}{m} N $ } & \scriptsize{Contact  \cite{Yang1967,Li:1995wr}}  \\ 
\scriptsize{$\exp(g|x_{i,i+1}|^{2})$} & \scriptsize{$ \displaystyle \frac{2\hbar^2 g}{m} \bigg[  N + 2g \sum_{i=1}^{N} |x_{i,i+1}|^{2} \bigg]  $} & \scriptsize{$ \displaystyle -\frac{4\hbar^2g^2}{m}  \sum_{i=1}^{N} x_{i-1,i} \ x_{i,i+1}  $} & \scriptsize{Coupled oscillators} \\ 

\scriptsize{$|\sinh{(x_{i,i+1}/\ell)}|^g$} & \scriptsize{$\displaystyle \frac{\hbar^2g}{m\ell^2} \bigg[ N + (g-1) \sum_{i=1}^{N} \coth^2{(x_{i,i+1}/\ell)} \bigg] $} & \scriptsize{$ \displaystyle -\frac{\hbar^2g^2}{m\ell^2} \sum_{i=1}^{N} \coth{(x_{i-1,i}/\ell)} \coth{(x_{i,i+1}/\ell) } $} & \scriptsize{Hyperbolic } \\ 

\end{tabular}

\caption{Quantum many-body systems associated with a cycle graph $C_N$ and ground-state wave function $\Phi_0=\prod_{i=1}^{N-1}f(x_{i,i+1})$ with  the particle index identifications $N+1\sim 1$ and $0\sim N$ for periodic boundary conditions. The three-body contribution generally becomes non-trivial even when, in a complete graph, it reduces to a constant or a two-body term; see Table \ref{KNmodels}.}
\label{CNmodels}
\end{table*}

Therefore, depending on the graph (whether it is $P_N$ or $C_N$), the Graph-Jastrow ansatz will be,
\begin{equation}
    \begin{split}
        \Phi_0^{(P_N)}(x_1,\cdot\cdot\cdot, x_N) & = \prod_{i=1}^{N-1} f_{i,i+1},  \\
        \Phi_0^{(C_N)}(x_1,\cdot\cdot\cdot, x_N) & =  \prod_{i=1}^{N} f_{i,(i+1) {\rm \ mod  \ }N}.
    \end{split}
\end{equation}

Now, using Eq. \eqref{V3Eq} we find that
\begin{eqnarray}\label{nn-Ham}
    \hat{H}_0^{(P_N)} & = & -\frac{\hbar^2}{2m} \sum_{i=1}^{N} \frac{\partial^2}{\partial x_i^2} + \hat{V}_2^{(P_N)} + \hat{V}_3^{(P_N)}, \\ 
    \hat{V}_2^{(P_N)} & = & +\frac{\hbar^2}{m} \sum_{i=1}^{N-1} 
    \frac{f''_{i,i+1}}{f_{i,i+1}}, \\
    \hat{V}_3^{(P_N)} &=& -\frac{\hbar^2}{m} \sum_{i=2}^{N-1} \frac{f'_{i-1,i}}{f_{i-1,i}}
    \frac{f'_{i,i+1}}{f_{i,i+1}}.
\end{eqnarray}

A specific instance of this family has been discussed in the literature and is known as the Jain-Khare model, with inverse-square pairwise interactions between nearest neighbors and three-body interactions restricted to next-nearest neighbors \cite{JainKhare99, Auberson01, Basu-Mallick01, Ezung05, Enciso05, Jain06}. This corresponds to the choice of the pair function $|x_{i,i+1}|^g$, which can be considered the nearest-neighbor truncation of the Calogero--Sutherland model (CSM).

Equation \eqref{nn-Ham} provides the complete family of one-dimensional models of the Jain-Khare type for an arbitrary choice of the pair function. Some relevant instances in this category are shown in Tables \ref{PNmodels} and \ref{CNmodels}.

Let us elaborate on the physical relevance of these models by explicitly considering the case with the pair correlation function $\exp(g|x_{ij}|)$, where the sum of $\hat{V}_2$ and $\hat{V}_3$ adds up to a constant and a two-body contact interaction between nearest neighbors. The latter is reminiscent of the Lieb-Liniger contact interactions that describe, e.g., ultracold gases in tight waveguides \cite{LL63, L63, Olshanii98}. As the interaction strength is arbitrary and possibly finite, this system describes distinguishable particles of equal mass, each interacting exclusively with two other particles (the neighbors in the graph representation) and exhibiting no interaction with the remaining particles.

% \vspace{-0.9 cm}

\section{$2r$ regular graph models: interactions with truncated range}
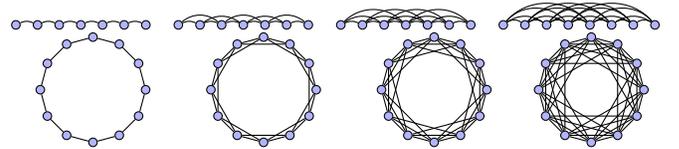
\begin{figure}[h]
\centering
\begin{tikzpicture}[>=Latex, scale=0.72]
% Define nodes
\node[circle, draw, fill=blue!30, inner sep=0.4mm] (n1) at (0,0) {};
\node[circle, draw, fill=blue!30, inner sep=0.4mm] (n2) at (0.4,0) {};
\node[circle, draw, fill=blue!30, inner sep=0.4mm] (n3) at (0.8,0) {};
\node[circle, draw, fill=blue!30, inner sep=0.4mm] (n4) at (1.2,0) {};
\node[circle, draw, fill=blue!30, inner sep=0.4mm] (n5) at (1.6,0) {};
\node[circle, draw, fill=blue!30, inner sep=0.4mm] (n6) at (2,0) {};
\node[circle, draw, fill=blue!30, inner sep=0.4mm] (n7) at (2.4,0) {};

% Define nodes
\node[circle, draw, fill=blue!30, inner sep=0.4mm] (n8) at (3,0) {};
\node[circle, draw, fill=blue!30, inner sep=0.4mm] (n9) at (3.4,0) {};
\node[circle, draw, fill=blue!30, inner sep=0.4mm] (n10) at (3.8,0) {};
\node[circle, draw, fill=blue!30, inner sep=0.4mm] (n11) at (4.2,0) {};
\node[circle, draw, fill=blue!30, inner sep=0.4mm] (n12) at (4.6,0) {};
\node[circle, draw, fill=blue!30, inner sep=0.4mm] (n13) at (5,0) {};
\node[circle, draw, fill=blue!30, inner sep=0.4mm] (n14) at (5.4,0) {};

% Define nodes
\node[circle, draw, fill=blue!30, inner sep=0.4mm] (n15) at (6,0) {};
\node[circle, draw, fill=blue!30, inner sep=0.4mm] (n16) at (6.4,0) {};
\node[circle, draw, fill=blue!30, inner sep=0.4mm] (n17) at (6.8,0) {};
\node[circle, draw, fill=blue!30, inner sep=0.4mm] (n18) at (7.2,0) {};
\node[circle, draw, fill=blue!30, inner sep=0.4mm] (n19) at (7.6,0) {};
\node[circle, draw, fill=blue!30, inner sep=0.4mm] (n20) at (8,0) {};
\node[circle, draw, fill=blue!30, inner sep=0.4mm] (n21) at (8.4,0) {};

% Define nodes
\node[circle, draw, fill=blue!30, inner sep=0.4mm] (n22) at (9,0) {};
\node[circle, draw, fill=blue!30, inner sep=0.4mm] (n23) at (9.4,0) {};
\node[circle, draw, fill=blue!30, inner sep=0.4mm] (n24) at (9.8,0) {};
\node[circle, draw, fill=blue!30, inner sep=0.4mm] (n25) at (10.2,0) {};
\node[circle, draw, fill=blue!30, inner sep=0.4mm] (n26) at (10.6,0) {};
\node[circle, draw, fill=blue!30, inner sep=0.4mm] (n27) at (11,0) {};
\node[circle, draw, fill=blue!30, inner sep=0.4mm] (n28) at (11.4,0) {};
\node[circle, draw, fill=blue!30, inner sep=0.4mm] (n29) at (11.8,0) {};

\foreach \i/\j/\bend in {1/2/25, 2/3/25, 3/4/25, 4/5/25, 5/6/25, 6/7/25, 8/9/25, 9/10/25, 10/11/25, 11/12/25, 12/13/25, 13/14/25,  8/10/40, 9/11/35, 10/12/40, 11/13/35, 12/14/40, 15/16/25, 16/17/25, 17/18/25, 18/19/25, 19/20/25, 20/21/25, 15/17/35, 16/18/35, 17/19/35, 18/20/35, 19/21/35,  15/18/45, 16/19/45, 17/20/45, 18/21/45, 22/23/35, 23/24/35, 24/25/35, 25/26/35, 26/27/35, 27/28/35, 28/29/35,  22/24/35, 23/25/35, 24/26/35, 25/27/35, 26/28/35, 27/29/35, 22/25/50, 23/26/50, 24/27/50, 25/28/50, 26/29/50, 22/26/50, 23/27/50, 24/28/50, 25/29/50}
{\draw[black, -, bend left=\bend] (n\i) to (n\j);}

\end{tikzpicture}
    \begin{minipage}{0.12\textwidth}
    \begin{tikzpicture}[scale=0.35, every node/.style={circle,draw,fill=blue!30,inner sep=0.4mm}]
    \def\n{12}
    \def\r{2cm}
    \foreach \i in {1,...,\n}{
        \node (V\i) at ({360/\n*(\i-1)}:\r) {};}
    \foreach \i in {1,...,\n}{
        \pgfmathtruncatemacro{\j}{mod(\i,\n)+1}
        \draw[black] (V\i) -- (V\j);}
\end{tikzpicture}
\end{minipage}
\begin{minipage}{0.12\textwidth}
    \begin{tikzpicture}[scale=0.35, every node/.style={circle,draw,fill=blue!30,inner sep=0.4mm}]
    \def\n{12}
    \def\r{2cm}
    \foreach \i in {1,...,\n}{
        \node (V\i) at ({360/\n*(\i-1)}:\r) {};}
    \foreach \i in {1,...,\n}{
        \pgfmathtruncatemacro{\j}{mod(\i,\n)+1}
        \draw[black] (V\i) -- (V\j);
        \pgfmathtruncatemacro{\j}{mod(\i+1,\n)+1} 
        \draw[black] (V\i) -- (V\j);}
\end{tikzpicture}
\end{minipage}
\begin{minipage}{0.12\textwidth}
    \begin{tikzpicture}[scale=0.35, every node/.style={circle,draw,fill=blue!30,inner sep=0.4mm}]
    \def\n{12}
    \def\r{2cm}
    \foreach \i in {1,...,\n}{
        \node (V\i) at ({360/\n*(\i-1)}:\r) {};}
    \foreach \i in {1,...,\n}{
        \pgfmathtruncatemacro{\j}{mod(\i,\n)+1}
        \draw[black] (V\i) -- (V\j);
        \pgfmathtruncatemacro{\j}{mod(\i+1,\n)+1} 
        \draw[black] (V\i) -- (V\j);
        \pgfmathtruncatemacro{\j}{mod(\i+2,\n)+1} 
        \draw[black] (V\i) -- (V\j);}
\end{tikzpicture}
\end{minipage}
    \begin{minipage}{0.1\textwidth}
    \begin{tikzpicture}[scale=0.35, every node/.style={circle,draw,fill=blue!30,inner sep=0.4mm}]
    \def\n{12}
    \def\r{2cm}e
    \foreach \i in {1,...,\n}{
        \node (V\i) at ({360/\n*(\i-1)}:\r) {};}
    \foreach \i in {1,...,\n}{
        \pgfmathtruncatemacro{\j}{mod(\i,\n)+1}
        \draw[black] (V\i) -- (V\j);
        \pgfmathtruncatemacro{\j}{mod(\i+1,\n)+1}
        \draw[black] (V\i) -- (V\j);
        \pgfmathtruncatemacro{\j}{mod(\i+2,\n)+1} 
        \draw[black] (V\i) -- (V\j);
        \pgfmathtruncatemacro{\j}{mod(\i+3,\n)+1} 
        \draw[black] (V\i) -- (V\j);}
\end{tikzpicture}
\end{minipage}
\caption{\label{FigureP6C6} Top: Regular graphs with $N=7,8$ and open boundary conditions and truncated pairwise interactions for $r=1,2,3,4$. Bottom: Circulant graphs with $N=12$ and periodic boundary conditions for $r=1,2,3,4$, the $2r$-regular graph interpolates between the cycle graph $C_{12}$ ($r=1$) and the complete graph $K_{12}$ ($r=6$).
}
\label{k_trunc_fig}
\end{figure}

Many-body models with a truncated range inverse-square interaction have recently been considered, interpolating between the Jain-Khare model and the Calogero-Sutherland model  \cite{Pittman17, Tummuru17}. This suggests the existence of a complete family of models with truncated range interactions (see Fig. \ref{k_trunc_fig}) involving other pair functions $f_{ij}$. That can be associated with the adjacency matrices

\begin{equation}\label{Achain}
\begin{aligned}
A^{(P_N)}_{ij} & = \sum_{k=1}^r \delta_{i,j+k} + \delta_{i,j-k} = \sum_{k=1}^r \delta_{|i-j|, k},  \\
A^{(C_N)}_{ij} & = \ \sum_{k=1}^r \delta_{i,(j+k) \mod{N}} + \delta_{i,(j-k) \mod{N}},
\end{aligned}
\end{equation}
where $r$ is the range of the interaction.  With periodic boundary conditions, the adjacency matrix describes a connected $(2r)$-regular graph in which each vertex has degree $2r$. Thus, the edge count is $|E|=r N$ and $p_2(\mathcal{G})=Nr(2r-1)$. Naturally, for $r=1$, one finds the $P_N$ and  $C_N$ models (with open and periodic boundary conditions, respectively), while for $r=(N-1)/2$ one recovers the $K_N$ models.

\begin{table*}[htbp] 
\setlength{\tabcolsep}{4pt}
\begin{tabular}{@{}llll@{}} 
\scriptsize{Pair function} & \scriptsize{Two-body potential: $\hat{V}_2$} & \scriptsize{Three-body potential: $\hat{V}_3$} & \scriptsize{Model/interactions} \\ 
\scriptsize{$f(x_{i,i+k})$} 
& \scriptsize{$\displaystyle \frac{\hbar^2}{m}   \sum_{k=1}^r \sum_{i=1}^{N-k}  
    \frac{f''_{i,i+k}}{f_{i,i+k}}$}
& \scriptsize{Eq. (\ref{V3T}) } 
& \scriptsize{Arbitrary interaction} \\ 

\scriptsize{$|x_{i,i+k}|^g$} & \scriptsize{$\displaystyle \frac{\hbar^2}{m} \sum_{k=1}^r \sum_{i=1}^{N-k}    \frac{g(g-1)}{|x_{i,i+k}|^2}$} & \scriptsize{Eq. (\ref{V3T})} & \scriptsize{TCSM \cite{Pittman17}} \\ 

\scriptsize{$\exp(g|x_{i,i+k}|)$} & \scriptsize{$\displaystyle \frac{\hbar^2g}{2m} \bigg[gr[2N-r-1] + 4 \sum_{k=1}^r \sum_{i=1}^{N-k} \delta(x_{i,i+k}) \bigg] $} & \scriptsize{Eq. (\ref{V3T}) } &   \\ 

\scriptsize{$\exp(g|x_{i,i+k}|^{2})$} & \scriptsize{$ \displaystyle \frac{\hbar^2g}{m} \bigg[r[2N-r-1] + 4g \sum_{k=1}^r \sum_{i=1}^{N-k} |x_{i,i+k}|^{2} \bigg]  $} & \scriptsize{Eq. (\ref{V3T})} &  \\ 

\scriptsize{$|\sinh{(x_{i,i+k}/\ell)}|^g $}& \scriptsize{$ \displaystyle \frac{\hbar^2g}{2m\ell^2} \bigg[ r[2N-r-1] + 2(g-1) \sum_{k=1}^r \sum_{i=1}^{N-k} \coth^2{(x_{i,i+k}/\ell)} \bigg]   $} & \scriptsize{Eq. (\ref{V3T})} &  \\

\end{tabular}
\caption{Quantum many-body systems associated with a $2r$-regular  graph and ground-state wave function $\Phi_0=\prod_{k=1}^r \prod_{i=1}^{N-k} f(x_{i,i+k})$  . The three-body contribution becomes non-trivial even in the cases where it reduces to a constant or a two-body term when the graph is complete; see Table \ref{KNmodels}.}
\label{2r-k_Truncated}
\end{table*}

We consider a ground-state with GJW form restricted to the $r$ nearest neighbors,
\beqa
\label{BJT}
\Phi_0(x_1,\dots,x_N) = \prod_{k=1}^r \prod_{i=1}^{N-k} f_{i,i+k}.
\eeqa

The parent Hamiltonian with the ground-state wave function (\ref{BJT}) takes the form of (\ref{Hgraph}), 
where the two-body potential involves interactions with a truncated range $r$
\begin{equation}\label{V2T}
     \begin{split}
         \hat{V}_2   = \frac{\hbar^2}{m} \sum_{k=1}^r  \sum_{i=1}^{N-k}     \frac{f''_{i,i+k}}{f_{i,i+k}} = \frac{\hbar^2}{m}  \sum_{\substack{i<j \\|i-j|\leq r}}^N \frac{f''_{ij}}{f_{ij}},
     \end{split}
\end{equation}
while the three-body term  simplifies to
\begin{eqnarray}
    \label{V3T}
    \hat{V}_3 & =& \frac{\hbar^2}{m} \bigg[ \sum_{k,q=1}^r \sum_{i=q+1}^{N-k}   \frac{f'_{i,i+k}}{f_{i,i+k}} \frac{f'_{i,i-q}}{f_{i,i-q}}  \\ & & + \sum_{k < q}^r\sum_{i=1}^{N-q}    \frac{f'_{i,i+k}}{f_{i,i+k}} \frac{f'_{i,i+q}}{f_{i,i+q}} + \sum_{k < q}^r \sum_{i=q+1}^{N}   \frac{f'_{i,i-k}}{f_{i,i-k}} \frac{f'_{i,i-q}}{f_{i,i-q}} \bigg]. \nonumber
\end{eqnarray}

Specific cases are listed in Table \ref{2r-k_Truncated}, where the form of the two-body potential $\hat{V}_2$ is given. All cases involve a nontrivial three-body term $\hat{V}_3$ that follows directly from Eq. \eqref{V3T}. For instance, for the truncated Calogero-Sutherland models (TCSM) \cite{Pittman17}, with $f(x_{i,j}) = |x_{i, j}|^g $, $f_{ij}'/f_{ij}=g/x_{ij}$. Similarly, choosing the pair function as a $k$-local exponential function, $f(x_{ij}) = \exp(g|x_{ij}|)$, yields $f_{ij}'/f_{ij}=g {\rm sgn}(x_{ij})$. As another choice, we consider the pair function given by a $k$-local Gaussian function, $f(x_{i,j}) = \exp(g|x_{i,j}|^{2})$, for which $f_{ij}'/f_{ij}=2gx_{ij}$. Finally, for the $k$-local hyperbolic function, $f(x_{i,j}) = |\sinh{(x_{i,j}/\ell)}|^g $, one can write down the explicit $\hat{V}_3$ term using $f_{ij}'/f_{ij}=(g/\ell)\coth{(x_{ij}/\ell)}$.

%%%%%%%%%%%%%%%%%%%%%% NEW SECTION %%%%%%%%%%%%%%%%%%%%

\section{Generating many-body quantum models by graph operations}

Beyond classifying models by graph family, our framework also allows the systematic construction of new ground-state-solvable systems by combining simpler graphs through elementary graph operations. Among them, the class of binary operations includes graph union, intersection, and join, as well as various graph products, which translate directly into composite many-body models with transparent physical interpretations. As the adjacency matrix enters directly into the ground-state wave function and the corresponding parent Hamiltonian, the construction of new models is natural using the tools of algebraic graph theory \cite{Godsil01}.

\subsection{Graph joins and mixtures}
Graph joins are a standard graph operation that combines two simpler graphs into a more complex one, which is otherwise difficult to study. Let $\mathcal{G}_1=(V_1, E_1)$ and $\mathcal{G}_2=(V_2, E_2)$ be two different graphs with $V_1 \cap V_2 = \varnothing $. The graph join $\mathcal{G}_1 \vee \mathcal{G}_2$ is obtained by taking the disjoint union $\mathcal{G}_1 \sqcup  \mathcal{G}_2$ and adding all possible edges between $V_1$ and $V_2$. Formally, we can write it as
\begin{eqnarray}
    E(\mathcal{G}_1 \vee \mathcal{G}_2) &=& E_1 \cup E_2 \cup \{ (v_1,v_2) : v_{1(2)} \in V_{1(2)} \}, \ \ \ \\ V(\mathcal{G}_1 \vee \mathcal{G}_2) &=& V_1 \cup V_2 .
\end{eqnarray}
Similarly, if $|V_1|=m, |V_2|=n$, the adjacency matrix can be written as
\begin{equation}
    A (\mathcal{G}_1 \vee \mathcal{G}_2) = \begin{pmatrix}
        A_1 & \mathbb{J}_{m\times n} \\ \mathbb{J}_{n\times m} & A_2
    \end{pmatrix}  . 
\end{equation}

\begin{figure}[h]
\centering
\begin{tikzpicture}[scale=1.15]

%% ================= LEFT GRAPH : K_{8,8} =================

%% Top partition (blue)
\node[circle, draw, fill=blue!30, inner sep=0.4mm] (n13)  at (0.8,0) {};
\node[circle, draw, fill=blue!30, inner sep=0.4mm] (n14)  at (1.2,0) {};
\node[circle, draw, fill=blue!30, inner sep=0.4mm] (n15)  at (1.6,0) {};
\node[circle, draw, fill=blue!30, inner sep=0.4mm] (n16)  at (2.0,0) {};
\node[circle, draw, fill=blue!30, inner sep=0.4mm] (n17)  at (2.4,0) {};
\node[circle, draw, fill=blue!30, inner sep=0.4mm] (n18)  at (2.8,0) {};
\node[circle, draw, fill=blue!30, inner sep=0.4mm] (n19)  at (3.2,0) {};
\node[circle, draw, fill=blue!30, inner sep=0.4mm] (n110) at (3.6,0) {};

%% Bottom partition (purple)
\node[circle, draw, fill=purple!50, inner sep=0.5mm] (n123) at (0.8,-1) {};
\node[circle, draw, fill=purple!50, inner sep=0.5mm] (n124) at (1.2,-1) {};
\node[circle, draw, fill=purple!50, inner sep=0.5mm] (n125) at (1.6,-1) {};
\node[circle, draw, fill=purple!50, inner sep=0.5mm] (n126) at (2.0,-1) {};
\node[circle, draw, fill=purple!50, inner sep=0.5mm] (n127) at (2.4,-1) {};
\node[circle, draw, fill=purple!50, inner sep=0.5mm] (n128) at (2.8,-1) {};
\node[circle, draw, fill=purple!50, inner sep=0.5mm] (n129) at (3.2,-1) {};
\node[circle, draw, fill=purple!50, inner sep=0.5mm] (n130) at (3.6,-1) {};

%% Edges of K_{8,8}
\foreach \i in {13,14,15,16,17,18,19,110}{
    \foreach \j in {123,124,125,126,127,128,129,130}{
        \draw[black,-] (n\i) -- (n\j);}}

%% Label (a)
\node at (2.2,-1.35) {\tiny{(a)}};

%% ================= RIGHT GRAPH : K_{8,5} =================

%% Top partition (blue)
\node[circle, draw, fill=blue!30, inner sep=0.4mm] (n22) at (5.2,0) {};
\node[circle, draw, fill=blue!30, inner sep=0.4mm] (n23) at (5.6,0) {};
\node[circle, draw, fill=blue!30, inner sep=0.4mm] (n24) at (6.0,0) {};
\node[circle, draw, fill=blue!30, inner sep=0.4mm] (n25) at (6.4,0) {};
\node[circle, draw, fill=blue!30, inner sep=0.4mm] (n26) at (6.8,0) {};
\node[circle, draw, fill=blue!30, inner sep=0.4mm] (n27) at (7.2,0) {};
\node[circle, draw, fill=blue!30, inner sep=0.4mm] (n28) at (7.6,0) {};
\node[circle, draw, fill=blue!30, inner sep=0.4mm] (n29) at (8.0,0) {};

%% Bottom partition (purple)
\node[circle, draw, fill=purple!50, inner sep=0.5mm] (n221) at (5.2,-1) {};
\node[circle, draw, fill=purple!50, inner sep=0.5mm] (n222) at (5.9,-1) {};
\node[circle, draw, fill=purple!50, inner sep=0.5mm] (n223) at (6.6,-1) {};
\node[circle, draw, fill=purple!50, inner sep=0.5mm] (n224) at (7.3,-1) {};
\node[circle, draw, fill=purple!50, inner sep=0.5mm] (n225) at (8.0,-1) {};

%% Edges of K_{8,5}
\foreach \i in {22,23,24,25,26,27,28,29}{
    \foreach \j in {221,222,223,224,225}{
        \draw[black,-] (n\i) -- (n\j);}}
\node at (6.6,-1.35) {\tiny{(b)}};
\end{tikzpicture}
\caption{Complete bipartite graphs: (a) $K_{8,8}$ and (b) $K_{8,5}$.}
\label{BipartiteGraph1}
\end{figure}
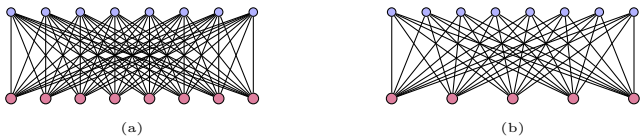

Some prominent examples of such graph joins are the complete bipartite graph $K_{m,n} = \overline{K_m} \vee \overline{K_n}$, which has $(m+n)$ vertices and an edge count of $|E|=mn$, with every vertex in $V_1$ connected to every vertex in $V_2$. The graphs $K_{8,8}$ and $K_{8,5}$ are shown in Fig. \ref{BipartiteGraph1}.  
\begin{table*}[htbp]
\begin{small}
\setlength{\tabcolsep}{4pt}
\begin{tabular}{@{}llll@{}} 
\scriptsize{Pair function} & \scriptsize{Two-body potential: $\hat{V}_2$} & \scriptsize{Three-body potential: $\hat{V}_3$} & \scriptsize{Model/interactions} \\ 
\scriptsize{$f(x_{1,j})$} 
& \scriptsize{$\displaystyle \frac{\hbar^2}{m}   \sum_{j=2}^N  
    \frac{f''_{1,j}}{f_{1,j}}$}
& \scriptsize{ $ \displaystyle \frac{\hbar^2}{m} \sum_{2\leq j < k}  \frac{f'_{1,j}}{f_{1,j}} \frac{f'_{1,k}}{f_{1,k}} $ } 
& \scriptsize{Arbitrary interaction} \\

\scriptsize{$|x_{1,j}|^g$} & \scriptsize{$\displaystyle \frac{\hbar^2}{m} \sum_{j=2}^N   \frac{g(g-1)}{|x_{1,j}|^2}$} & \scriptsize{ $ \displaystyle \frac{\hbar^2g^2}{m} \sum_{2\leq j < k} \frac{1}{x_{1,j} \ x_{1,k}} $ } & \\ 

\scriptsize{$\exp(g|x_{1,j}|)$} & \scriptsize{$\displaystyle \frac{\hbar^2g}{m} \bigg[  g(N-1) + 2 \sum_{j=2}^N \delta(x_{1,j}) \bigg]  $} & \scriptsize{ $ \displaystyle \frac{\hbar^2g^2}{m} \sum_{2\leq j < k}  {\rm sgn}(x_{1,j}) \ {\rm sgn}(x_{1,k})  $ } &  \\ 

\scriptsize{$\exp(g|x_{1,j}|^{2})$} & \scriptsize{$ \displaystyle \frac{2\hbar^2g}{m} \bigg[ (N-1) + 2g \sum_{j=2}^N |x_{1,j}|^2 \bigg]   $} & \scriptsize{  $\displaystyle \frac{4\hbar^2g^2}{m} \sum_{2\leq j < k} x_{1,j} \ x_{1,k} $  } &  \\ 

\scriptsize{$|\sinh{(x_{1,j}/\ell)}|^g $}& \scriptsize{$ \displaystyle \frac{\hbar^2g}{m\ell^2} \bigg[ (N-1) + (g-1) \sum_{j=2}^N \coth^2{(x_{1,j}/\ell)}   \bigg]   $} & \scriptsize{  $\displaystyle \frac{\hbar^2g^2}{m\ell^2} \sum_{2\leq j < k} \coth{(x_{1,j}/\ell)} \ \coth{(x_{1,k}/\ell)} $   } &  \\

\end{tabular}
\end{small}
\caption{Quantum many-body systems associated with a star graph and ground-state wave function $\Phi_0=\prod_{j=2}^N f(x_{1,j})$. The three-body contribution becomes a restricted three-body term even when, in a complete graph, it reduces to a constant or a two-body term; see Table \ref{KNmodels}.}
\label{star_graph_table}
\end{table*}

Bipartite graphs are natural for describing binary mixtures of particles. In particular, consider a system composed of two species $A$ and $B$. Assume there are $N_A$ particles of type $A$ and $N_B$ particles of type $B$. A bipartite graph $K_{N_A, N_B}$ is natural to describe mixtures in which there are no $AA$ or $BB$ interactions, but there are $AB$ interactions. Arguably, this setting is rather restricted, as one may generally want to consider $AA$ and $B$ interactions, with or without local restrictions. 

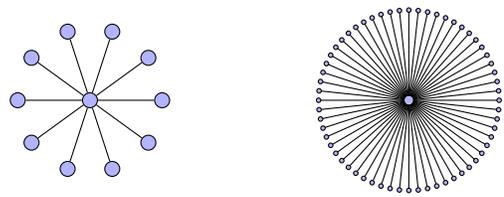
\begin{figure}[h]
\centering

% ---------- GRAPH 1 ----------
\begin{minipage}{0.23\textwidth}
\centering
\begin{tikzpicture}[scale=0.6]
    \def\N{10}
    \def\R{1.6}

    \node[circle,draw,fill=blue!30,inner sep=2pt] (0) at (0,0) {};

    \foreach \i in {1,...,\N} {
        \node[circle,draw,fill=blue!30,inner sep=2pt] (n\i)
           at ({\R*cos(360/\N*\i)}, {\R*sin(360/\N*\i)}) {};
        \draw[black] (0) -- (n\i);}
\end{tikzpicture}
\end{minipage}
%
% ---------- GRAPH 2 ----------
\begin{minipage}{0.23\textwidth}
\centering
\begin{tikzpicture}[scale=0.6]
    \def\N{60}
    \def\R{2}

    \node[circle,fill=blue!30,inner sep=1pt] (0) at (0,0) {};

    \foreach \i in {1,...,\N} {
        \coordinate (p\i)
           at ({\R*cos(360/\N*\i)}, {\R*sin(360/\N*\i)});
        \draw[black,thin] (0) -- (p\i);}

    \foreach \i in {1,...,\N} {
        \node[circle,draw,fill=blue!30,inner sep=0.6pt]
          at ({\R*cos(360/\N*\i)}, {\R*sin(360/\N*\i)}) {};}
\end{tikzpicture}
\end{minipage}
\caption{{Star graphs} with $N=10,60$; describing a quantum many-body system of central-spin or impurity system.}
\label{star_graph_fig}
\end{figure}

We will consider another instructive example: the star graph $S_{n} = K_1 \vee \overline{K_n}$; see Fig. \ref{star_graph_fig}. Many-body systems with pair correlations described by a star graph often arise in models of decoherence with an explicit description of the environmental degrees of freedom, as in the central spin model. A star graph is a tree in which one vertex has degree $N-1$, while the remaining $(N-1)$ vertices have degree 1. A star graph is also a complete bipartite graph $K_{1, N-1}$. The ground-state GJW in such a system is of the form
\beqa
\label{BJstar}
\Phi_0(x_1,\dots,x_N)=\prod_{j=2}^Nf(x_{1,j})=\prod_{j=2}^N f_{1,j}  ,
\eeqa

\noindent where the particle at position $x_1$ occupies the central vertex of the connectivity graph, with adjacency matrix $A_{ij}=\delta_{i,1}+\delta_{1,j}-2\delta_{i,1}\delta_{1,j}$.
Physically, this corresponds to a single particle embedded in a surrounding environment of distinguishable particles, as in impurity and polaron models, with the important distinction that the latter case often involves indistinguishable particles. Many-body states with pair correlations described by a star graph are the ground states of the Hamiltonian with

\begin{eqnarray}
    \hat{H}_0 &=& -\frac{\hbar^2}{2m} \sum_{i=1}^N \frac{\partial^2}{\partial x_i^2} + \hat{V}_2 + \hat{V}_3 , \\
\label{V2star}
    \hat{V}_2 &=& \frac{\hbar^2}{m}  \sum_{j=2}^N\frac{f''_{1,j}}{f_{1,j}} , \\
     \label{V3star}
    \hat{V}_3 &=& -\frac{\hbar^2}{m} \sum_{2\leq j <k}^N \frac{f'_{j,1}}{f_{j,1}}  \frac{f'_{1,k}}{f_{1,k}}.
\end{eqnarray}

$\hat{V}_2$ and $\hat{V}_3$ for different choices of $f_{ij}$ can be found in Table \ref{star_graph_table} . One can also write the graph joins as $\mathcal{G}_1 \vee \mathcal{G}_2 = \overline{\overline{\mathcal{G}_1} \sqcup \overline{\mathcal{G}_2}  }$. Quite similar to the star graphs, one can also write wheel graphs as the graph join, $W_{N+1} =K_1 \vee C_N$, which we discuss in the next section.
\subsection{Graph products}
\begin{table*}
\setlength{\tabcolsep}{4pt}
\begin{tabular}{@{}lll@{}} 

\textbf{\scriptsize{Graph product type}} & \textbf{\scriptsize{Edge condition description}} & \textbf{\scriptsize{Adjacency matrix}} \\ 

\scriptsize{Cartesian : $\mathcal{G}_1 \square \mathcal{G}_2$} 
& \scriptsize{$v_1$ \& $v_2$ are adjacent if $g_1=g_2$ \& $\{h_1,h_2\}\in E(\mathcal{G}_2)$, or vice-versa} 
& \scriptsize{$A_{\mathcal{G}_1} \otimes \mathbb{I}_{\mathcal{G}_2} + \mathbb{I}_{\mathcal{G}_1} \otimes A_{\mathcal{G}_2}$} \\ 

\scriptsize{Lexicographic : $\mathcal{G}_1 \circ \mathcal{G}_2$} 
& \scriptsize{$v_1$ \& $v_2$ are adjacent if $\{g_1,g_2\}\in E(\mathcal{G}_1)$ or $g_1=g_2$ \& $\{h_1,h_2\}\in E(\mathcal{G}_2)$} 
& \scriptsize{$A_{\mathcal{G}_1} \otimes \mathbb{J}_{\mathcal{G}_2} + \mathbb{I}_{\mathcal{G}_1} \otimes A_{\mathcal{G}_2}$} \\ 

\scriptsize{Tensor/Direct : $\mathcal{G}_1 \times \mathcal{G}_2$} 
& \scriptsize{$v_1$ \& $v_2$ are adjacent if $\{g_1,g_2\}\in E(\mathcal{G}_1)$ \& $\{h_1,h_2\}\in E(\mathcal{G}_2)$} 
& \scriptsize{$A_{\mathcal{G}_1} \otimes A_{\mathcal{G}_2}$} \\ 

\scriptsize{Strong : $\mathcal{G}_1 \boxtimes \mathcal{G}_2$} 
& \scriptsize{$v_1$ \& $v_2$ are adjacent if Cartesian or Tensor condition is satisfied} 
& \scriptsize{$A_{\mathcal{G}_1} \otimes \mathbb{I}_{\mathcal{G}_2} + \mathbb{I}_{\mathcal{G}_1} \otimes A_{\mathcal{G}_2} + A_{\mathcal{G}_1} \otimes A_{\mathcal{G}_2} $} \\ 

\scriptsize{Corona : $\mathcal{G}_1 \odot \mathcal{G}_2$} 
& \scriptsize{Attach a copy of $\mathcal{G}_2$ to each vertex of $\mathcal{G}_1$} 
& \scriptsize{Not simple Kronecker} \\

\end{tabular}
\caption{Summary of common graph products for two graphs $\mathcal{G}_1$ and $\mathcal{G}_2$, including their adjacency matrix representation. Here,  $v_1=(g_1,h_1)$ and $v_2=(g_2,h_2)$ are vertices of graph $\mathcal{G}_1$ and $\mathcal{G}_2$, such that $\{g_k\} \in \mathcal{G}_1$ and $\{h_q\} \in \mathcal{G}_2$. $\mathbb{I}$ is the identity matrix, and $\mathbb{J}$ is the all-ones matrix.}
\label{Grph_prod}
\end{table*}
Our discussion of star and wheel graphs motivates considering more complex graphs with $N-1$ vertices as a model of an environment $\mathcal{G}_\mathcal{E}$ in which all nodes interact with an additional central particle, in analogy with central-spin models. As it turns out, the connectivity graph of such a composite many-body system is given by a graph operation, the corona product $K_1\odot\mathcal{G}_\mathcal{E}$. More generally, in the description of a composite system including an environment and a system of interest, the connectivity graph is described by the graph join $\mathcal{G}_\mathcal{E}\vee \mathcal{G}_\mathcal{S}$  obtained by the union graph $\mathcal{G}_\mathcal{E}\cup\mathcal{G}_\mathcal{S}$ of the environment with that of the system,  with all the edges that join vertices of the first graph to vertices of the second. The Hilbert space of the composite system is the tensor product $\mathcal{H}_{\mathcal{G}_\mathcal{E}}\otimes\mathcal{H}_{\mathcal{G}_\mathcal{S}}$. An alternative construction of more complex many-body quantum systems from simple ones relies on graph products, of which one can distinguish different kinds; see Table \ref{Grph_prod}. Let us consider two graphs $\mathcal{G}_1$ and $\mathcal{G}_2$,  with adjacency matrices $A_1$ and $A_2$ and vertex counts $|V_1|$ and $|V_2|$, respectively.  The corresponding many-body quantum systems have Hilbert spaces $\mathcal{H}_{\mathcal{G}_1}$ and $\mathcal{H}_{\mathcal{G}_2}$. The Cartesian product of the two graphs  is a new graph $\mathcal{G}_1\Box\mathcal{G}_2$ with $|V_1|\times|V_2|$ vertices and adjacency matrix $A(\mathcal{G}_1\Box\mathcal{G}_2)=A(\mathcal{G}_2\Box\mathcal{G}_1)$
\beqa
A(\mathcal{G}_1\Box\mathcal{G}_2)=A(\mathcal{G}_1)\otimes\mathbb{I}_{\mathcal{G}_2}+\mathbb{I}_{\mathcal{G}_1}\otimes A(\mathcal{G}_2) .
\eeqa

\begin{figure}[h]
    \centering
    \begin{tikzpicture}
        \draw[purple, thick] (0,0)--(0.5,0);
        \draw[purple, thick] (0.5,0)--(1,0);
        \draw[purple, thick] (1,0)--(1.5,0);
        \draw[purple, thick] (1.5,0)--(2,0);
        \draw[purple, thick] (2,0)--(2.5,0);
        \draw[purple, thick] (2.5,0)--(3,0);
        %%%%%%%%%%%%%%%%%%%%%%%%%%%%%%%%%%%%%%%%
        \draw[purple, thick] (0,0.5)--(0.5,0.5);
        \draw[purple, thick] (0.5,0.5)--(1,0.5);
        \draw[purple, thick] (1,0.5)--(1.5,0.5);
        \draw[purple, thick] (1.5,0.5)--(2,0.5);
        \draw[purple, thick] (2,0.5)--(2.5,0.5);
        \draw[purple, thick] (2.5,0.5)--(3,0.5);
        %%%%%%%%%%%%%%%%%%%%%%%%%%%%%%%%%%%%%%%%%
        \draw[thick](0,0)--(0,0.5);
        \draw[thick](0.5,0)--(0.5,0.5);
        \draw[thick](1,0)--(1,0.5);
        \draw[thick](1.5,0)--(1.5,0.5);
        \draw[thick](2,0)--(2,0.5);
        \draw[thick](2.5,0)--(2.5,0.5);
        \draw[thick](3,0)--(3,0.5);
        %%%%%%%%%%%%%%%%
        \filldraw[fill=blue!50,inner sep=1pt] (0,0) circle (1.5 pt);
        \filldraw[fill=blue!50,inner sep=1pt] (0.5,0) circle (1.5 pt);
        \filldraw[fill=blue!50,inner sep=1pt] (1,0) circle (1.5 pt);
        \filldraw[fill=blue!50,inner sep=1pt] (1.5,0) circle (1.5 pt);
        \filldraw[fill=blue!50,inner sep=1pt] (2,0) circle (1.5 pt);
        \filldraw[fill=blue!50,inner sep=1pt] (2.5,0) circle (1.5 pt);
        \filldraw[fill=blue!50,inner sep=1pt] (3,0) circle (1.5 pt);
        \filldraw[fill=blue!50,inner sep=1pt] (0,0.5) circle (1.5 pt);
        \filldraw[fill=blue!50,inner sep=1pt] (0.5,0.5) circle (1.5 pt);
        \filldraw[fill=blue!50,inner sep=1pt] (1,0.5) circle (1.5 pt);
        \filldraw[fill=blue!50,inner sep=1pt] (1.5,0.5) circle (1.5 pt);
        \filldraw[fill=blue!50,inner sep=1pt] (2,0.5) circle (1.5 pt);
        \filldraw[fill=blue!50,inner sep=1pt] (2.5,0.5) circle (1.5 pt);
        \filldraw[fill=blue!50,inner sep=1pt] (3,0.5) circle (1.5 pt);
        %%%%%%%%%%%%%%%%%%%%%%%%%%%%%%%%%%%%%%%%%%%%%%%%%%%%%%%%%%%%%%%%%%
        \draw[thick, decorate, decoration={coil, amplitude=0.9pt, segment length=1.6pt}](0,-0.55)--(0.5,-0.55);
        \draw[thick, decorate, decoration={coil, amplitude=0.9pt, segment length=1.6pt}](1,-0.55)--(1.5,-0.55);
        \draw[thick, decorate, decoration={coil, amplitude=0.9pt, segment length=1.6pt}](2,-0.55)--(2.5,-0.55);
        \draw[thick, decorate, decoration={coil, amplitude=0.9pt, segment length=1.6pt}](3,-0.55)--(3.5,-0.55);
        \draw[thick, decorate, decoration={coil, amplitude=0.9pt, segment length=1.6pt}](4,-0.55)--(4.5,-0.55);
        \draw[thick, decorate, decoration={coil, amplitude=0.9pt, segment length=1.6pt}](5,-0.55)--(5.5,-0.55);
        \draw[thick, decorate, decoration={coil, amplitude=0.9pt, segment length=1.6pt}](6,-0.55)--(6.5,-0.55);

        %%%%%%%%%%%%%%%%%%%%%%%%%%%%%%%%%%%%%%%%%%%%%%%%%%%%%%%%%%%%%%%%%%
        \node[circle, draw, fill=blue!30, inner sep=0.01pt] (n1) at (0,-0.55) {\tiny{A}};
        \node[circle, draw, fill=blue!30, inner sep=0.01pt] (n2) at (0.5,-0.55) {\tiny{B}};
        \node[circle, draw, fill=blue!30, inner sep=0.01pt] (n3) at (1,-0.55) {\tiny{A}};
        \node[circle, draw, fill=blue!30, inner sep=0.01pt] (n4) at (1.5,-0.55) {\tiny{B}};
        \node[circle, draw, fill=blue!30, inner sep=0.01pt] (n5) at (2,-0.55) {\tiny{A}};
        \node[circle, draw, fill=blue!30, inner sep=0.01pt] (n6) at (2.5,-0.55) {\tiny{B}};
        \node[circle, draw, fill=blue!30, inner sep=0.01pt] (n7) at (3,-0.55) {\tiny{A}};
        \node[circle, draw, fill=blue!30, inner sep=0.01pt] (n8) at (3.5,-0.55) {\tiny{B}};
        \node[circle, draw, fill=blue!30, inner sep=0.01pt] (n9) at (4,-0.55) {\tiny{A}};
        \node[circle, draw, fill=blue!30, inner sep=0.01pt] (n10) at (4.5,-0.55) {\tiny{B}};
        \node[circle, draw, fill=blue!30, inner sep=0.01pt] (n11) at (5,-0.55) {\tiny{A}};
        \node[circle, draw, fill=blue!30, inner sep=0.01pt] (n12) at (5.5,-0.55) {\tiny{B}};
        \node[circle, draw, fill=blue!30, inner sep=0.01pt] (n13) at (6,-0.55) {\tiny{A}};
        \node[circle, draw, fill=blue!30, inner sep=0.01pt] (n14) at (6.5,-0.55) {\tiny{B}};
   %%%%%%%%%%%%%%%%%%%%%%%%%
   \foreach \i/\j/\bend in {1/3/25, 2/4/-25, 3/5/25, 4/6/-25, 5/7/25, 6/8/-25, 7/9/25, 8/10/-25, 9/11/25, 10/12/-25, 11/13/25, 12/14/-25}
{\draw[purple, thick, -, bend left=\bend] (n\i) to (n\j);}
    \end{tikzpicture}
    \caption{Ladder graph $L_7=P_7 \square P_2$ (top) and flattened ladder graph (bottom) realized as a 1D model with two particles bound per unit cell, which can be visualized as $x_i = (x_i^A,x_i^B)$.}
    \label{ladder_grph}
\end{figure}
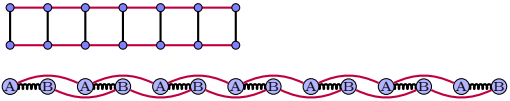

The Hilbert space of the new many-body system on the graph $\mathcal{G}_1\Box\mathcal{G}_2$  is thus $\mathcal{H}_{\mathcal{G}_1\Box\mathcal{G}_2}=\mathcal{H}_{\mathcal{G}_1}\otimes\mathcal{H}_{\mathcal{G}_2}$. A relevant example in condensed matter is the ladder graph (see Fig. \ref{ladder_grph}), $L_N=P_N\Box P_2$, with $|V|=2N$ and $|E|=3N-2$. 

The adjacency matrix for the ladder graph, compressed to 1D as two composite particles bound to form a single entity, can be written as
\begin{equation}
    A_{(i,\alpha),(j,\beta)} = \delta_{|i-j|,1}  \delta_{\alpha,\beta} + \delta_{i,j} \delta_{\alpha,\Bar{\beta}},
\end{equation}
where $i\in\{1,\cdot\cdot\cdot,N\}$, $\alpha \in \{A,B\}$, and $\Bar{A}=B , \Bar{B}=A$. Using this, we can write the ground-state GJW,
\begin{equation}
    \begin{split}
        \Phi_0(\{x_k^A, x_k^B\}) & =  \prod_{(i,\alpha) < (j,\beta) } \bigg[ f_{i,j}^{(\alpha,\beta)} \bigg]^{A_{(i,\alpha),(j,\beta)}}   \\
        & = \prod_{i=1}^N f(x_i^A - x_i^B)  \prod_{\alpha=A,B} \prod_{i=1}^N f_{i,i+1}^{\alpha}  , 
    \end{split}
\end{equation}
where $f_{i,j}^{(\alpha,\beta)} = f(x_i^{\alpha} - x_j^{\beta})$, $f_{i,i}^{(\alpha,\beta)}=f_i^{(\alpha,\beta)}$, and  $f^{(\alpha,\alpha)}_{i,j}=f^{\alpha}_{i,j}$. Using this GJW ansatz, we find the parent Hamiltonian 
\begin{eqnarray}
     \hat{H}_0 & = & \sum_{\alpha=A,B}\hat{H}_0^{\alpha} = \hat{H}_0^A + \hat{H}_0^B  , \\ \hat{H}_0^{\alpha} & = & -\frac{\hbar^2}{2m} \sum_{i} \frac{\partial^2}{\partial (x_i^{\alpha})^2} +  \hat{V}_{\rm int}^{\alpha} + \hat{V}_{2}^{\alpha} + \hat{V}_{2 \rm L}^{\alpha} + \hat{V}_3^{\alpha}   ,  \nonumber 
\end{eqnarray}
where the extra two-body term $V_{\rm int}$ accounts for the nearest-neighbor interaction between $A$ and $B$ particles. Explicitly, we can write all the multi body terms as [$\eta_{\alpha}=+(-)1 $ for $\alpha= A(B)$]
\begin{eqnarray}
\hat{V}_{\rm int}^{\alpha} & = & \frac{\hbar^2}{2m} \sum_{i=1}^N \frac{f''(x_i^A - x_i^B)}{f(x_i^A - x_i^B)}  , \\
\hat{V}_2^{\alpha} & = & \frac{\hbar^2}{m} \sum_{i} \frac{f''(x_i^{\alpha}-x_{i+1}^{\alpha})}{f(x_i^{\alpha}-x_{i+1}^{\alpha})}  , \\
    \hat{V}_{2\rm L}^{\alpha} & = & \frac{\hbar^2}{m} \sum_{i}   \frac{f'^{(A,B)}_i}{f^{(A,B)}_i} \bigg[\frac{f'^{\alpha}_{i-1,i}}{f^{\alpha}_{i-1,i}} - \frac{f'^{\alpha}_{i,i+1}}{f^{\alpha}_{i,i+1}} \bigg] \eta_{\alpha}  , \\
    \hat{V}_3^{\alpha} & = & -\frac{\hbar^2}{m} \sum_{i} \frac{f'^{\alpha}_{i-1,i}}{f^{\alpha}_{i-1,i}} \frac{f'^{\alpha}_{i,i+1}}{f^{\alpha}_{i,i+1}}  .
\end{eqnarray}
\begin{table*}
\setlength{\tabcolsep}{4pt}
\begin{tabular}{@{}lll@{}} 

\textbf{\scriptsize{Standard graph}} & \textbf{\scriptsize{Product type}} & \textbf{\scriptsize{Product structure}} \\ 

\scriptsize{Grid/lattice $P_{m,n}$} 
& \scriptsize{Cartesian} 
& \scriptsize{$P_m \square P_n$} \\ 

\scriptsize{Hypercube $Q_n$} 
& \scriptsize{Cartesian} 
& \scriptsize{$K_2^{\square n} = K_2 \square K_2 \square \cdot\cdot\cdot \square K_2 \ \ $ ($n$ times)} \\ 

\scriptsize{Complete bipartite $K_{m,n}$} 
& \scriptsize{Lexicographic} 
& \scriptsize{$E_m \circ K_n $} \\ 

\scriptsize{Prism $Y_n$} 
& \scriptsize{Cartesian} 
& \scriptsize{$C_n \square P_2 $} \\ 

\scriptsize{Wheel $W_{n+1}$  } 
& \scriptsize{Corona/join} 
& \scriptsize{$K_1 \odot C_n$ or $K_1 \vee C_n $} \\

\end{tabular}
\caption{Summary of standard graphs produced via graph products of two or more simple graphs.}
\label{Grph_prod_ex}
\end{table*}
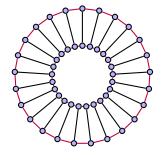
\begin{figure}[h]
    \centering
    \begin{tikzpicture}[scale=0.9,
    every node/.style={circle, draw, fill=blue!30,inner sep=0.7pt},
     path/.style={purple!70}]

\def\R{1}
\def\r{0.45}
\def\n{25}  

\foreach \i in {1,...,\n}
  \node (O\i) at ({90+360/\n*(\i-1)}:\R) {};

\foreach \i [evaluate=\i as \j using {int(mod(\i,\n)+1)}] in {1,...,\n}
  \draw[purple] (O\i)--(O\j);

\foreach \i in {1,...,\n}
  \node (I\i) at ({90+360/\n*(\i-1)}:\r) {};

\foreach \i [evaluate=\i as \j using {int(mod(\i,\n)+1)}] in {1,...,\n}
  \draw[purple] (I\i)--(I\j);

\foreach \i in {1,...,\n}
  \draw (O\i)--(I\i);
\end{tikzpicture}
    \caption{Quantum many-body systems represented in the form of a prism graph $Y_{25}=C_{25} \square P_2$. See Table \ref{Grph_prod_ex}. }
    \label{prism_grph}
\end{figure}

Similarly,  prism graphs, also known as circular ladder graphs, are associated with the Cartesian product of a cycle graph and a $2$-path, $Y_N = CL_N = C_N \Box P_2$, see Fig. \ref{prism_grph}. A finite $N\times M$ lattice graph can be obtained as $P_N\Box P_M$. 
In particular, for $r=3$, one finds cubic graphs $Q_N$, such as the Petersen graph ($N=10$). The adjacency matrix for prism graphs is
\begin{equation}
    \begin{split}
        A_{ij}   = &  \sum_{k=1}^N \bigg[ \delta_{i,k} \delta_{j, (k \  {\rm mod} \  N)+1} + \delta_{i,k} \delta_{j,k+N} \\ & \ \ \ \ \  +  \delta_{i,N+k} \delta_{j, (k \  {\rm mod} \  N)+N+1}  \bigg].
    \end{split}
\end{equation}

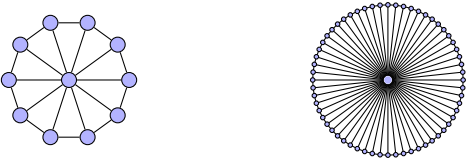
\begin{figure}[h!]
\centering
% ---------- GRAPH 3 ----------
\begin{minipage}{0.23\textwidth}
\centering
\begin{tikzpicture}[scale=0.5]
    \def\N{10}
    \def\R{1.6}

    \node[circle,draw,fill=blue!30,inner sep=2pt] (0) at (0,0) {};

    \foreach \i in {1,...,\N} {
        \node[circle,draw,fill=blue!30,inner sep=2pt] (n\i)
           at ({\R*cos(360/\N*\i)}, {\R*sin(360/\N*\i)}) {};}

    % cycle edges
    \foreach \i in {1,...,\N} {
        \pgfmathtruncatemacro{\j}{mod(\i,\N)+1}
        \draw[black] (n\i) -- (n\j);}

    % spokes
    \foreach \i in {1,...,\N} {
        \draw[black] (0) -- (n\i);}
\end{tikzpicture}
\end{minipage}
%
% ---------- GRAPH 4 ----------
\begin{minipage}{0.23\textwidth}
\centering
\begin{tikzpicture}[scale=0.5]
    \def\N{60}
    \def\R{2}

    \node[circle,fill=blue!30,inner sep=1pt] (0) at (0,0) {};

    \foreach \i in {1,...,\N} {
        \coordinate (n\i)
           at ({\R*cos(360/\N*\i)}, {\R*sin(360/\N*\i)});
        \draw[black,thin] (0) -- (n\i);}

    % cycle edges
    \foreach \i in {1,...,\N} {
        \pgfmathtruncatemacro{\j}{mod(\i,\N)+1}
        \draw[black,thin] (n\i) -- (n\j);}

    \foreach \i in {1,...,\N} {
        \node[circle,draw,fill=blue!30,inner sep=0.6pt]
          at ({\R*cos(360/\N*\i)}, {\R*sin(360/\N*\i)}) {};}
\end{tikzpicture}
\end{minipage}
\caption{Wheel graphs with $N=10,60$ vertices, which is formed from the graph operation, $W_{N+1}=K_1 \odot C_N$.}
\label{wheel_graph_fig}
\end{figure}

The many-body systems associated with a star graph are involved in an environment of non-interacting particles. The simplest generalization is to include nearest-neighbor interactions between the particles in the environment, while keeping their pairwise interactions with the central particle. This leads to considering pair correlations characterized by a wheel graph $W_{N+1}$ (see Fig. \ref{wheel_graph_fig}), obtained from a cycle graph $C_{N}$ by adding a vertex that is a hub and is connected to all the vertices of $C_{N}$. As a result, $|E|=2N$. The adjacency matrix has elements
\beqa
\begin{split}
A_{ij}= & \ \delta_{i,1}+\delta_{1,j}-2\delta_{i,1}\delta_{1,j}+\delta_{i,j+1}+\delta_{i,j-1} \\ & + \delta_{i,2}\delta_{N,j}+\delta_{i,N}\delta_{2,j} - \delta_{i,2} \delta_{1,j} - \delta_{i,1}\delta_{2} .
\end{split}
\eeqa

The ground-state GJW for the wheel graph reads 
\begin{equation}
    \begin{split}
        \Phi_0  & = \prod_{k=1}^{n-1}  f_{k, k+1}   f_{k+n, k+n+1 }  f_{k,k+n} f_{n,1} f_{2n, n+1} f_{n,2n }.
    \end{split}
\end{equation}

\begin{figure}[h]
    \centering
    \begin{tikzpicture}
        \draw[purple, thick] (0,0)--(0.5,0);
        \draw[purple, thick] (0.5,0)--(1,0);
        \draw[purple, thick] (1,0)--(1.5,0);
        \draw[purple, thick] (1.5,0)--(2,0);
        \draw[purple, thick] (2,0)--(2.5,0);
        \draw[purple, thick] (2.5,0)--(3,0);
        %%%%%%%%%%%%%%%%%%%%%%%%%%%%%%%%%%%%%%%%
        \draw[purple, thick] (0,0.5)--(0.5,0.5);
        \draw[purple, thick] (0.5,0.5)--(1,0.5);
        \draw[purple, thick] (1,0.5)--(1.5,0.5);
        \draw[purple, thick] (1.5,0.5)--(2,0.5);
        \draw[purple, thick] (2,0.5)--(2.5,0.5);
        \draw[purple, thick] (2.5,0.5)--(3,0.5);
        %%%%%%%%%%%%%%%%%%%%%%%%%%%%%%%%%%%%%%%%%
        \draw[thick](0,0)--(0,0.5);
        \draw[thick](0.5,0)--(0.5,0.5);
        \draw[thick](1,0)--(1,0.5);
        \draw[thick](1.5,0)--(1.5,0.5);
        \draw[thick](2,0)--(2,0.5);
        \draw[thick](2.5,0)--(2.5,0.5);
        \draw[thick](3,0)--(3,0.5);
        %%%%%%%%%%%%%%%%%%%%%%%%%%%%%%%%%%%%%%%%%%%%%%%%%%%%%%%%%%%%%%%%%%
        \draw[thick, teal](0,0)--(0.5,0.5);
        \draw[thick, teal](0.5,0)--(1,0.5);
        \draw[thick, teal](1,0)--(1.5,0.5);
        \draw[thick, teal](1.5,0)--(2,0.5);
        \draw[thick, teal](2,0)--(2.5,0.5);
        \draw[thick, teal](2.5,0)--(3,0.5);
        \draw[thick, orange](0,0.5)--(0.5,0);
        \draw[thick, orange](0.5,0.5)--(1,0);
        \draw[thick, orange](1,0.5)--(1.5,0);
        \draw[thick, orange](1.5,0.5)--(2,0);
        \draw[thick, orange](2,0.5)--(2.5,0);
        \draw[thick, orange](2.5,0.5)--(3,0);
        %%%%%%%%%%%%%%
        \filldraw[fill=blue!50,inner sep=1pt] (0,0) circle (1.5 pt);
        \filldraw[fill=blue!50,inner sep=1pt] (0.5,0) circle (1.5 pt);
        \filldraw[fill=blue!50,inner sep=1pt] (1,0) circle (1.5 pt);
        \filldraw[fill=blue!50,inner sep=1pt] (1.5,0) circle (1.5 pt);
        \filldraw[fill=blue!50,inner sep=1pt] (2,0) circle (1.5 pt);
        \filldraw[fill=blue!50,inner sep=1pt] (2.5,0) circle (1.5 pt);
        \filldraw[fill=blue!50,inner sep=1pt] (3,0) circle (1.5 pt);
        \filldraw[fill=blue!50,inner sep=1pt] (0,0.5) circle (1.5 pt);
        \filldraw[fill=blue!50,inner sep=1pt] (0.5,0.5) circle (1.5 pt);
        \filldraw[fill=blue!50,inner sep=1pt] (1,0.5) circle (1.5 pt);
        \filldraw[fill=blue!50,inner sep=1pt] (1.5,0.5) circle (1.5 pt);
        \filldraw[fill=blue!50,inner sep=1pt] (2,0.5) circle (1.5 pt);
        \filldraw[fill=blue!50,inner sep=1pt] (2.5,0.5) circle (1.5 pt);
        \filldraw[fill=blue!50,inner sep=1pt] (3,0.5) circle (1.5 pt);
        %%%%%%%%%%%%%%%%%%%%%%%%%%%%%%%%%%%%%%%%%%%%%%%%%%%%%%%%%%%%%%%%%%
        \draw[thick, decorate, decoration={coil, amplitude=0.9pt, segment length=1.6pt}](0,-0.55)--(0.5,-0.55);
        \draw[thick, decorate, decoration={coil, amplitude=0.9pt, segment length=1.6pt}](1,-0.55)--(1.5,-0.55);
        \draw[thick, decorate, decoration={coil, amplitude=0.9pt, segment length=1.6pt}](2,-0.55)--(2.5,-0.55);
        \draw[thick, decorate, decoration={coil, amplitude=0.9pt, segment length=1.6pt}](3,-0.55)--(3.5,-0.55);
        \draw[thick, decorate, decoration={coil, amplitude=0.9pt, segment length=1.6pt}](4,-0.55)--(4.5,-0.55);
        \draw[thick, decorate, decoration={coil, amplitude=0.9pt, segment length=1.6pt}](5,-0.55)--(5.5,-0.55);
        \draw[thick, decorate, decoration={coil, amplitude=0.9pt, segment length=1.6pt}](6,-0.55)--(6.5,-0.55);
        %%%%%%%%%%%%%%%%%%%%%%%%%%%%%%%%%%%%%%%%%%%%%%%%%%%%%%%%%%%%%%%%%%
        \node[circle, draw, fill=blue!30, inner sep=0.01pt] (n1) at (0,-0.55) {\tiny{A}};
        \node[circle, draw, fill=blue!30, inner sep=0.01pt] (n2) at (0.5,-0.55) {\tiny{B}};
        \node[circle, draw, fill=blue!30, inner sep=0.01pt] (n3) at (1,-0.55) {\tiny{A}};
        \node[circle, draw, fill=blue!30, inner sep=0.01pt] (n4) at (1.5,-0.55) {\tiny{B}};
        \node[circle, draw, fill=blue!30, inner sep=0.01pt] (n5) at (2,-0.55) {\tiny{A}};
        \node[circle, draw, fill=blue!30, inner sep=0.01pt] (n6) at (2.5,-0.55) {\tiny{B}};
        \node[circle, draw, fill=blue!30, inner sep=0.01pt] (n7) at (3,-0.55) {\tiny{A}};
        \node[circle, draw, fill=blue!30, inner sep=0.01pt] (n8) at (3.5,-0.55) {\tiny{B}};
        \node[circle, draw, fill=blue!30, inner sep=0.01pt] (n9) at (4,-0.55) {\tiny{A}};
        \node[circle, draw, fill=blue!30, inner sep=0.01pt] (n10) at (4.5,-0.55) {\tiny{B}};
        \node[circle, draw, fill=blue!30, inner sep=0.01pt] (n11) at (5,-0.55) {\tiny{A}};
        \node[circle, draw, fill=blue!30, inner sep=0.01pt] (n12) at (5.5,-0.55) {\tiny{B}};
        \node[circle, draw, fill=blue!30, inner sep=0.01pt] (n13) at (6,-0.55) {\tiny{A}};
        \node[circle, draw, fill=blue!30, inner sep=0.01pt] (n14) at (6.5,-0.55) {\tiny{B}};
   %%%%%%%%%%%%%%%%%%%%%%%%%
   \foreach \i/\j/\bend in {1/3/25, 2/4/-25, 3/5/25, 4/6/-25, 5/7/25, 6/8/-25, 7/9/25, 8/10/-25, 9/11/25, 10/12/-25, 11/13/25, 12/14/-25}
{\draw[purple, thick, -, bend left=\bend] (n\i) to (n\j);}
\foreach \i/\j/\bend in {1/4/35, 2/5/-35, 3/6/35, 4/7/-35, 5/8/35, 6/9/-35, 7/10/35, 8/11/-35, 9/12/35, 10/13/-35, 11/14/35}
{\draw[orange, thick, -, bend left=\bend] (n\i) to (n\j);}
\foreach \i/\j/\bend in {2/3/0, 4/5/0, 6/7/0, 8/9/0, 10/11/0, 12/13/0}
{\draw[teal, thick, -, bend left=\bend] (n\i) to (n\j);}
    \end{tikzpicture}
    \caption{Top: The Creutz ladder graph $P_7 \circ P_2 \cong P_7 \boxtimes P_2$, comes from a theorem $\mathcal{G}_1 \circ \mathcal{G}_2 \cong \mathcal{G}_1 \boxtimes \mathcal{G}_2 $ iff $\mathcal{G}_2$ is a complete graph \cite{Theorem_mondal2025}. Bottom: Flattened Creutz ladder to visualize a more realistic 1D model with two particles bound per unit cell. It differs from the model shown in Fig. \ref{ladder_grph}, by including the cross-particle interactions in the nearest neighbor.}
    \label{Corss_ladder_grph}
\end{figure}
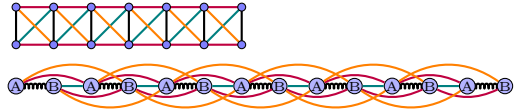

A few other relevant graph operations are the strong product and the lexicographic product. The strong product graph $\mathcal{G}_1\boxtimes\mathcal{G}_2$ and the lexicographic product $\mathcal{G}_1\circ \mathcal{G}_2$ of $\mathcal{G}_1$ and $\mathcal{G}_2$ have a vertex count of $|V_{\mathcal{G}_1}||V_{\mathcal{G}_2}|$, and their adjacency matrices reads as follows:
\begin{eqnarray}
    A(\mathcal{G}_1\boxtimes \mathcal{G}_2) &=& A_{\mathcal{G}_1} \otimes \mathbb{I}_{\mathcal{G}_2} + \mathbb{I}_{\mathcal{G}_1} \otimes A_{\mathcal{G}_2} + A_{\mathcal{G}_1} \otimes A_{\mathcal{G}_2} 
 , \ \ \  {} \\
    A(\mathcal{G}_1\circ \mathcal{G}_2) &=& A_{\mathcal{G}_1} \otimes \mathbb{J}_{\mathcal{G}_2} + \mathbb{I}_{\mathcal{G}_1} \otimes A_{\mathcal{G}_2} ,
\end{eqnarray}
where $\mathbb{J}$ is the all-ones matrix. Consequently, the many-body Hilbert space is $\mathcal{H}_{\mathcal{G}_1\times\mathcal{G}_2}=\mathcal{H}_{\mathcal{G}_1}\otimes\mathcal{H}_{\mathcal{G}_2}$. A relevant example to demonstrate the following two graph operations is the celebrated Creutz ladder \cite{Creutz99}, shown in Fig. \ref{Corss_ladder_grph}.

The adjacency matrix for the Creutz ladder graph, compressed to 1D as two composite particles bound to form a single entity, can be written as
\begin{equation}
    A_{(i,\alpha),(j,\beta)} = \delta_{|i-j|,1}  \delta_{\alpha,\beta} + \delta_{|i-j|,1} \delta_{\alpha,\Bar{\beta}} + \delta_{i,j} \delta_{\alpha,\Bar{\beta}},
\end{equation}
where, $i\in\{1,\cdot\cdot\cdot,N\}$, $\alpha \in \{A,B\}$, and $\Bar{A}=B , \Bar{B}=A$.
% \newline 

Similarly, there are other known graph products available in the literature %[suitable citation?]%, 
a non-exhaustive set of such products with descriptions is given in the Table \ref{Grph_prod}. We emphasize that the graph structure is not necessarily reflected by pinning the particles in coordinate space as one would do for particles sitting on a lattice. Rather, its adjacency matrix governs the ground-state correlations and particle interactions.

\section{Discussion}

Before closing, we consider several extensions of our study. The formulation of parent Hamiltonians of graph Jastrow wave functions presented focuses on the graph structure of pairwise interparticle interactions. The coordinate of each particle takes values in the Euclidean space $\vec{r}_i\in\mathbb{R}^D$ and is unconstrained by the graph features, except for the possible occurrence of hard-core interparticle interactions. In condensed matter systems, it is often of interest to restrict the motion of particles to a given spatial distribution. One possible way to do so is to use a structured external potential, in which particles can move as in, e.g., ultracold atoms in optical lattices and optical tweezers. In other applications, it is preferable to pin down the particles at a given location in space, e.g., as in the description of quantum solids, lattices of oscillators, etc. In Monte Carlo studies, such pinning is described by replacing the Jastrow ansatz by a Nosanow-Jastrow wave function, in which the location of each particle is determined \cite{Cazorla07, Cazorla09, delcampo20}. At this level of description,  particles are distinguished by their location, e.g., on a lattice or, more generally, a topological graph. This may be pursued using topological graph theory, which studies the embedding of a graph in a given space. 

Another research direction is that of disordered systems. Random many-body interactions are central to the study of spin glasses, many-body localization, the Sachdev--Ye--Kitaev model, and quantum chaos more broadly \cite{Binder1986, Vojta2019}, and the present framework accommodates them naturally. Consider the ground-states of the generalized GJW form
\beqa
\Phi_0(x_1,\dots,x_N)=\prod_{i<j}e^{p_{ij}\ln f_{ij}},
\eeqa
with $p_{ij}\in\mathbb{R}$, to accommodate weighted graphs or random graphs. In this case, the parent Hamiltonian takes the form
\beqa
\hat{H}_0=-\frac{\hbar^2}{2m}\sum_{i=1}^{N}\frac{\partial^2}{\partial x_i^2} +\hat{V}_2+ \hat{V} _3,
\eeqa
with the two-body and three-body potentials given by (assuming $p_{ij} = p_{ji}$)
\begin{align}
\hat{V}_2 &= \frac{\hbar^2}{m} \sum_{i<j}
\left[
p_{ij} \frac{f''_{ij}}{f_{ij}}
+ p_{ij}(p_{ij}-1)
\bigg(\frac{f'_{ij}}{f_{ij}}\bigg)^2
\right],
\label{V2Eqp}  \\
\hat{V}_3 &= \frac{\hbar^2}{m} \sum_{i<j<k}
\Bigg[
p_{ij}p_{ik}\frac{f'_{ij}f'_{ik}}{f_{ij}f_{ik}}
- p_{ij}p_{jk}\frac{f'_{ij}f'_{jk}}{f_{ij}f_{jk}}
\notag\\
&\qquad
+ p_{ik}p_{jk}\frac{f'_{ik}f'_{jk}}{f_{ik}f_{jk}}
\Bigg].
\label{V3Eqp}
\end{align}

This resembles a multi species many-body system in which the interaction strength varies for different pairs and trios of particles. Such a possibility has been studied in the literature, and we may find important generalizations. For instance, the generalization of the CSM to multiple species with varying interactions was presented in both one \cite{Meljanac03} and higher spatial dimensions \cite{Meljanac04}. In random graphs, one considers a distribution of graphs with a probability density function, e.g., for a given graph. In other words, one can consider the probability distribution for a given adjacency matrix ${\rm Pr}(A)$ and study, e.g., the average  Hamiltonian.
\beqa
\int d\mu(A){\rm Pr}(A) \hat{H}_0(A).
\eeqa

Ensemble averages over pure states lead to mixed states represented by density matrices, motivating the construction
\beqa
\int d\mu(A){\rm Pr}(A) \Phi_0(x_1,\dots,x_N;A)\Phi_0^*(x_1',\dots,x_N';A).\nonumber \\
\eeqa

Extensions to higher spatial dimensions and to random or regular graph ensembles may reveal universality classes controlled by connectivity rather than by geometry.

More broadly, the graph formulation suggests that qualitative features of many-body systems, such as the prevalence and structure of three-body terms or the emergence of effective locality, can be organized by standard graph invariants, such as the degree distribution, number of 2-paths, clustering, and the spectral properties of $A$. Identifying graph families that preserve extended sets of conserved quantities and characterizing excitation spectra are particularly promising open problems. Specifically, beyond constructing a tower of excited states associated with the center of mass, one may wonder whether the parent Hamiltonians of GJW may be quasi exactly solvable \cite{Ushveridze1994, Turbiner11}.

Yet another frontier is that of nonequilibrium phenomena, where the use of a time-dependent Jastrow ansatz \cite{Sutherland98, Yang2025} facilitates numerous studies, including quantum decay \cite{Jukic08, delcampo11, delcampo16}, Loschmidt echoes \cite{Ares18}, dynamical fermionization and bosonization \cite{Minguzzi2005, Buljan2008, delcampo08, Dupays23}, and quantum speed limits \cite{delcampo18, Fogarty20}.

\section{Conclusions}

We have introduced a graph-based generalization of the Jastrow ansatz for many-body systems of distinguishable continuous-variable particles. In this ``Graph-Jastrow'' construction, the ground-state wave function is a product of pair correlators over the edge set of a graph, and the adjacency matrix determines which correlations are present. The approach provides a continuum-variable counterpart of spin models on graphs, with explicit breaking of permutation symmetry for generic connectivity.

A central structural result is that the corresponding parent Hamiltonian contains (i) two-body interactions supported on graph edges and (ii) three-body interactions supported on length-2 paths of the graph. These terms arise from the action of the kinetic-energy operator on an edge-factorized wave function and therefore encode the local combinatorics of the underlying connectivity. In the complete-graph limit, the construction recovers standard Bijl--Jastrow quantum fluids and known integrable models; for restricted connectivity, it reproduces and extends truncated and symmetry-broken families discussed in the literature.

More generally, by organizing models according to graph families and graph operations (joins and products), we have charted a landscape of integrable and ground-state solvable systems for which the Hamiltonian, ground state, and ground-state energy are obtained in closed form. In one dimension, the framework also provides an alternative notion of locality: interactions are truncated by adjacency rather than by spatial range.

This approach can be readily generalized to other exchange statistics, including fermions, composite fermions \cite{Jain2007}, and anyons \cite{Kundu99, Girardeau06}. It can also be extended to account for particles with internal degrees of freedom \cite{Yang1967, Kawakami1993, Kawakami93, GirardeauOlshanii04}, from which spin models can be derived by freezing, e.g., as done in the Haldane-Shastry models \cite{Haldane88, Shastry88, Polychronakos93, Sutherland04, Polychronakos06}.  

Open directions include understanding excitation spectra and correlations beyond the ground state, establishing large-$N$ limits for sparse and dense graph sequences, and identifying graph conditions under which additional conserved quantities survive. Extensions to weighted and random graph ensembles, as well as to spatially embedded or pinned-particle variants, offer further opportunities to connect the present construction with experimental platforms and with established paradigms in many-body physics. 

\section{Acknowledgments}
It is a pleasure to acknowledge discussions with Errol D. G. Drummond, \'I\~nigo L. Egusquiza, Andr\'as Grabarits, Maxim Olchanyi, and Jing Yang. This work is supported by the  Luxembourg National Research Fund under Grant No. C-PRIDE/23/18691647/QUANCOM.

\bibliography{MBQGraphs_lib}

@article{Jukic08,
  title = {Free expansion of a Lieb-Liniger gas: Asymptotic form of the wave functions},
  author = {Juki\ifmmode \acute{c}\else \'{c}\fi{}, D. and Pezer, R. and Gasenzer, T. and Buljan, H.},
  journal = {Phys. Rev. A},
  volume = {78},
  issue = {5},
  pages = {053602},
  numpages = {9},
  year = {2008},
  month = {Nov},
  publisher = {American Physical Society},
  doi = {10.1103/PhysRevA.78.053602},
  url = {https://link.aps.org/doi/10.1103/PhysRevA.78.053602}
}

@article{Dupays23,
  title = {Tailoring dynamical fermionization: Delta-kick cooling of a Tonks-Girardeau gas},
  author = {Dupays, L\'eonce and Yang, Jing and del Campo, Adolfo},
  journal = {Phys. Rev. A},
  volume = {107},
  issue = {5},
  pages = {L051302},
  numpages = {6},
  year = {2023},
  month = {May},
  publisher = {American Physical Society},
  doi = {10.1103/PhysRevA.107.L051302},
  url = {https://link.aps.org/doi/10.1103/PhysRevA.107.L051302}
}

@article{Buljan2008,
  title = {Fermi-Bose Transformation for the Time-Dependent Lieb-Liniger Gas},
  author = {Buljan, H. and Pezer, R. and Gasenzer, T.},
  journal = {Phys. Rev. Lett.},
  volume = {100},
  issue = {8},
  pages = {080406},
  numpages = {4},
  year = {2008},
  month = {Feb},
  publisher = {American Physical Society},
  doi = {10.1103/PhysRevLett.100.080406},
  url = {https://link.aps.org/doi/10.1103/PhysRevLett.100.080406}
}

@article{delcampo11,
  title = {Long-time behavior of many-particle quantum decay},
  author = {del Campo, A.},
  journal = {Phys. Rev. A},
  volume = {84},
  issue = {1},
  pages = {012113},
  numpages = {6},
  year = {2011},
  month = {Jul},
  publisher = {American Physical Society},
  doi = {10.1103/PhysRevA.84.012113},
  url = {https://link.aps.org/doi/10.1103/PhysRevA.84.012113}
}

@article{Minguzzi2005,
  title = {Exact Coherent States of a Harmonically Confined Tonks-Girardeau Gas},
  author = {Minguzzi, A. and Gangardt, D. M.},
  journal = {Phys. Rev. Lett.},
  volume = {94},
  issue = {24},
  pages = {240404},
  numpages = {4},
  year = {2005},
  month = {Jun},
  publisher = {American Physical Society},
  doi = {10.1103/PhysRevLett.94.240404},
  url = {https://link.aps.org/doi/10.1103/PhysRevLett.94.240404}
}

@article{Nosanow66,
  title = {Theory of Quantum Crystals},
  author = {Nosanow, L. H.},
  journal = {Phys. Rev.},
  volume = {146},
  issue = {1},
  pages = {120--133},
  numpages = {0},
  year = {1966},
  month = {Jun},
  publisher = {American Physical Society},
  doi = {10.1103/PhysRev.146.120},
  url = {https://link.aps.org/doi/10.1103/PhysRev.146.120}
}

@book{KuramotoKato09, 
place={Cambridge}, 
title={Dynamics of One-Dimensional Quantum Systems: Inverse-Square Interaction Models}, 
DOI={10.1017/CBO9780511596827}, 
publisher={Cambridge University Press}, 
author={Kuramoto, Yoshio and Kato, Yusuke}, year={2009}}

@article{Calogero69,
    author = {Calogero, F.},
    title = "{Solution of a Three‐Body Problem in One Dimension}",
    journal = {Journal of Mathematical Physics},
    volume = {10},
    number = {12},
    pages = {2191-2196},
    year = {2003},
    month = {11},
    issn = {0022-2488},
    doi = {10.1063/1.1664820},
    url = {https://doi.org/10.1063/1.1664820}
}

@article{Moser75,
title = {Three integrable Hamiltonian systems connected with isospectral deformations},
journal = {Advances in Mathematics},
volume = {16},
number = {2},
pages = {197-220},
year = {1975},
issn = {0001-8708},
doi = {https://doi.org/10.1016/0001-8708(75)90151-6},
url = {https://www.sciencedirect.com/science/article/pii/0001870875901516},
author = {J. Moser}
}

@article{Olshanetsky81,
title = {Classical integrable finite-dimensional systems related to Lie algebras},
journal = {Physics Reports},
volume = {71},
number = {5},
pages = {313-400},
year = {1981},
issn = {0370-1573},
doi = {https://doi.org/10.1016/0370-1573(81)90023-5},
url = {https://www.sciencedirect.com/science/article/pii/0370157381900235},
author = {M. A. Olshanetsky and A. M. Perelomov}
}

@article{Olshanetsky83,
title = {Quantum integrable systems related to lie algebras},
journal = {Physics Reports},
volume = {94},
number = {6},
pages = {313-404},
year = {1983},
issn = {0370-1573},
doi = {https://doi.org/10.1016/0370-1573(83)90018-2},
url = {https://www.sciencedirect.com/science/article/pii/0370157383900182},
author = {M. A. Olshanetsky and A. M. Perelomov}
}

@article{Turbiner11,
   title={From Quantum AN(Calogero) to H4(Rational) Model},
   ISSN={1815-0659},
   url={http://dx.doi.org/10.3842/SIGMA.2011.071},
   DOI={10.3842/sigma.2011.071},
   journal={Symmetry, Integrability and Geometry: Methods and Applications},
   publisher={SIGMA (Symmetry, Integrability and Geometry: Methods and Application)},
   author={Turbiner, Alexander V.},
   year={2011},
   month=jul }

@article{Harshman17,
  title = {Integrable Families of Hard-Core Particles with Unequal Masses in a One-Dimensional Harmonic Trap},
  author = {Harshman, N. L. and Olshanii, Maxim and Dehkharghani, A. S. and Volosniev, A. G. and Jackson, Steven Glenn and Zinner, N. T.},
  journal = {Phys. Rev. X},
  volume = {7},
  issue = {4},
  pages = {041001},
  numpages = {14},
  year = {2017},
  month = {Oct},
  publisher = {American Physical Society},
  doi = {10.1103/PhysRevX.7.041001},
  url = {https://link.aps.org/doi/10.1103/PhysRevX.7.041001}
}

@article{Beau20,
  title = {Exactly Solvable System of One-Dimensional Trapped Bosons with Short- and Long-Range Interactions},
  author = {Beau, M. and Pittman, S. M. and Astrakharchik, G. E. and del Campo, A.},
  journal = {Phys. Rev. Lett.},
  volume = {125},
  issue = {22},
  pages = {220602},
  numpages = {7},
  year = {2020},
  month = {Nov},
  publisher = {American Physical Society},
  doi = {10.1103/PhysRevLett.125.220602},
  url = {https://link.aps.org/doi/10.1103/PhysRevLett.125.220602}
}

@article{GirardeauOlshanii04,
  title = {Theory of spinor Fermi and Bose gases in tight atom waveguides},
  author = {Girardeau, M. D. and Olshanii, M.},
  journal = {Phys. Rev. A},
  volume = {70},
  issue = {2},
  pages = {023608},
  numpages = {4},
  year = {2004},
  month = {Aug},
  publisher = {American Physical Society},
  doi = {10.1103/PhysRevA.70.023608},
  url = {https://link.aps.org/doi/10.1103/PhysRevA.70.023608}
}

@Article{Beau21,
	title={{Parent Hamiltonians of Jastrow wavefunctions}},
	author={Mathieu Beau and Adolfo del Campo},
	journal={SciPost Phys. Core},
	volume={4},
	pages={030},
	year={2021},
	publisher={SciPost},
	doi={10.21468/SciPostPhysCore.4.4.030},
	url={https://scipost.org/10.21468/SciPostPhysCore.4.4.030},
}

@article{Yang22,
  title = {One-Dimensional Quantum Systems with Ground State of Jastrow Form Are Integrable},
  author = {Yang, Jing and del Campo, Adolfo},
  journal = {Phys. Rev. Lett.},
  volume = {129},
  issue = {15},
  pages = {150601},
  numpages = {6},
  year = {2022},
  month = {Oct},
  publisher = {American Physical Society},
  doi = {10.1103/PhysRevLett.129.150601},
  url = {https://link.aps.org/doi/10.1103/PhysRevLett.129.150601}
}

@article{mackel2022quantum,
  doi = {10.22331/q-2023-12-20-1211},
  url = {https://doi.org/10.22331/q-2023-12-20-1211},
  title = {Quantum {A}lchemy and {U}niversal {O}rthogonality {C}atastrophe in {O}ne-{D}imensional {A}nyons},
  author = {Mackel, Naim E. and Yang, Jing and Campo, Adolfo del},
  journal = {{Quantum}},
  issn = {2521-327X},
  publisher = {{Verein zur F{\"{o}}rderung des Open Access Publizierens in den Quantenwissenschaften}},
  volume = {7},
  pages = {1211},
  month = dec,
  year = {2023}
}

@book{Godsil01,
author = {Chris Godsil and Gordon F. Royle},
publisher = {Springer},
title = {Algebraic Graph Theory},
year = {2001},
doi = {10.1007/978-1-4613-0163-9},
address = {New York}
}

@Article{Jain06,
author={Jain, Sudhir R.},
title={Random matrix theories and exactly solvable models},
journal={Czechoslovak Journal of Physics},
year={2006},
month={Sep},
day={01},
volume={56},
number={9},
pages={1021-1032},
issn={1572-9486},
doi={10.1007/s10582-006-0397-7},
url={https://doi.org/10.1007/s10582-006-0397-7}
}

@article{Meljanac03,
title = "A multispecies Calogero model",
journal = "Physics Letters B",
volume = "573",
pages = "202 - 208",
year = "2003",
issn = "0370-2693",
doi = "https://doi.org/10.1016/j.physletb.2003.08.029",
url = "http://www.sciencedirect.com/science/article/pii/S0370269303012929",
author = "S Meljanac and M Milekovi\'c and A Samsarov"
}

@article{Meljanac04,
title = "Generalized Calogero model in arbitrary dimensions",
journal = "Physics Letters B",
volume = "594",
number = "1",
pages = "241 - 246",
year = "2004",
issn = "0370-2693",
doi = "https://doi.org/10.1016/j.physletb.2004.05.034",
url = "http://www.sciencedirect.com/science/article/pii/S0370269304007816",
author = "S. Meljanac and M. Milekovi\'c and A. Samsarov"
}

@ARTICLE{delcampo20,
  title = {Exact ground states of quantum many-body systems under confinement},
  author = {del Campo, Adolfo},
  journal = {Phys. Rev. Research},
  volume = {2},
  issue = {4},
  pages = {043114},
  numpages = {7},
  year = {2020},
  month = {Oct},
  publisher = {American Physical Society},
  doi = {10.1103/PhysRevResearch.2.043114},
  url = {https://link.aps.org/doi/10.1103/PhysRevResearch.2.043114}
}

@article{Cazorla07,
	doi = {10.1088/0953-8984/20/01/015223},
	url = {https://doi.org/10.1088%2F0953-8984%2F20%2F01%2F015223},
	year = 2007,
	month = {dec},
	publisher = {{IOP} Publishing},
	volume = {20},
	number = {1},
	pages = {015223},
	author = {C Cazorla and J Boronat},
	title = {Zero-temperature equation of state of solid4He at low and high pressures},
	journal = {Journal of Physics: Condensed Matter}
}

@article{Dhar10,
author = { Abhishek   Dhar },
title = {Heat transport in low-dimensional systems},
journal = {Advances in Physics},
volume = {57},
number = {5},
pages = {457-537},
year  = {2008},
publisher = {Taylor & Francis},
doi = {10.1080/00018730802538522},
URL = {
        https://doi.org/10.1080/00018730802538522
}
}

@article{Eisert10,
  title = {Colloquium: Area laws for the entanglement entropy},
  author = {Eisert, J. and Cramer, M. and Plenio, M. B.},
  journal = {Rev. Mod. Phys.},
  volume = {82},
  issue = {1},
  pages = {277--306},
  numpages = {0},
  year = {2010},
  month = {Feb},
  publisher = {American Physical Society},
  doi = {10.1103/RevModPhys.82.277},
  url = {https://link.aps.org/doi/10.1103/RevModPhys.82.277}
}

@article{Cazorla09,
	doi = {10.1088/1367-2630/11/1/013047},
	url = {https://doi.org/10.1088%2F1367-2630%2F11%2F1%2F013047},
	year = 2009,
	month = {jan},
	publisher = {{IOP} Publishing},
	volume = {11},
	number = {1},
	pages = {013047},
	author = {C Cazorla and G E Astrakharchik and J Casulleras and J Boronat},
	title = {Bose{\textendash}Einstein quantum statistics and the ground state of solid4He},
	journal = {New Journal of Physics}
}

@article{LL63,
  title = {Exact Analysis of an Interacting Bose Gas. I. The General Solution and the Ground State},
  author = {Lieb, Elliott H. and Liniger, Werner},
  journal = {Phys. Rev.},
  volume = {130},
  issue = {4},
  pages = {1605--1616},
  numpages = {0},
  year = {1963},
  month = {May},
  publisher = {American Physical Society},
  doi = {10.1103/PhysRev.130.1605},
  url = {https://link.aps.org/doi/10.1103/PhysRev.130.1605}
}

@article{Kawakami93,
author = {Kawakami ,Norio},
title = {Renormalized Harmonic-Oscillator Description of   Confined Electron Systems with Inverse-Square Interaction},
journal = {Journal of the Physical Society of Japan},
volume = {62},
number = {12},
pages = {4163-4166},
year = {1993},
doi = {10.1143/JPSJ.62.4163},
URL = {https://doi.org/10.1143/JPSJ.62.4163}}

@article{Enciso05,
title = "Solvable scalar and spin models with near-neighbors interactions",
journal = "Physics Letters B",
volume = "605",
number = "1",
pages = "214 - 222",
year = "2005",
issn = "0370-2693",
doi = "https://doi.org/10.1016/j.physletb.2004.11.031",
url = "http://www.sciencedirect.com/science/article/pii/S0370269304016004",
author = "A. Enciso and F. Finkel and A. Gonz\'alez-L\'opez and M.A. Rodr\'iguez"
}

@article{Batchelor06,
  title = {One-Dimensional Interacting Anyon Gas: Low-Energy Properties and Haldane Exclusion Statistics},
  author = {Batchelor, M. T. and Guan, X.-W. and Oelkers, N.},
  journal = {Phys. Rev. Lett.},
  volume = {96},
  issue = {21},
  pages = {210402},
  numpages = {4},
  year = {2006},
  month = {Jun},
  publisher = {American Physical Society},
  doi = {10.1103/PhysRevLett.96.210402},
  url = {https://link.aps.org/doi/10.1103/PhysRevLett.96.210402}
}

@article{delcampo18,
  title = {Probing Quantum Speed Limits with Ultracold Gases},
  author = {del Campo, Adolfo},
  journal = {Phys. Rev. Lett.},
  volume = {126},
  issue = {18},
  pages = {180603},
  numpages = {6},
  year = {2021},
  month = {May},
  publisher = {American Physical Society},
  doi = {10.1103/PhysRevLett.126.180603},
  url = {https://link.aps.org/doi/10.1103/PhysRevLett.126.180603}
}

@article{Tummuru17,
title = "Truncated Calogero–Sutherland models on a circle",
journal = "Physics Letters A",
volume = "381",
number = "47",
pages = "3917 - 3920",
year = "2017",
issn = "0375-9601",
doi = "https://doi.org/10.1016/j.physleta.2017.10.007",
url = "http://www.sciencedirect.com/science/article/pii/S0375960116309677",
author = "Tarun R. Tummuru and Sudhir R. Jain and Avinash Khare"
}

@article{Girardeau06,
  title = {Anyon-Fermion Mapping and Applications to Ultracold Gases in Tight Waveguides},
  author = {Girardeau, M. D.},
  journal = {Phys. Rev. Lett.},
  volume = {97},
  issue = {10},
  pages = {100402},
  numpages = {4},
  year = {2006},
  month = {Sep},
  publisher = {American Physical Society},
  doi = {10.1103/PhysRevLett.97.100402},
  url = {https://link.aps.org/doi/10.1103/PhysRevLett.97.100402}
}

@article{Kundu99,
  title = {Exact Solution of Double $\ensuremath{\delta}$ Function Bose Gas through an Interacting Anyon Gas},
  author = {Kundu, Anjan},
  journal = {Phys. Rev. Lett.},
  volume = {83},
  issue = {7},
  pages = {1275--1278},
  numpages = {0},
  year = {1999},
  month = {Aug},
  publisher = {American Physical Society},
  doi = {10.1103/PhysRevLett.83.1275},
  url = {https://link.aps.org/doi/10.1103/PhysRevLett.83.1275}
}

@article{L63,
  title = {Exact Analysis of an Interacting Bose Gas. II. The Excitation Spectrum},
  author = {Lieb, Elliott H.},
  journal = {Phys. Rev.},
  volume = {130},
  issue = {4},
  pages = {1616--1624},
  numpages = {0},
  year = {1963},
  month = {May},
  publisher = {American Physical Society},
  doi = {10.1103/PhysRev.130.1616},
  url = {https://link.aps.org/doi/10.1103/PhysRev.130.1616}
}

@article{Calogero71,
author = {Calogero, F. },
title = {Solution of the One‐Dimensional N‐Body Problems with Quadratic and/or Inversely Quadratic Pair Potentials},
journal = {Journal of Mathematical Physics},
volume = {12},
number = {3},
pages = {419-436},
year = {1971},
doi = {10.1063/1.1665604},
URL = {https://doi.org/10.1063/1.1665604}
}

@Article{Bethe31,
author={Bethe, H.},
title={Zur Theorie der Metalle},
journal={Zeitschrift f{\"u}r Physik},
year={1931},
month={Mar},
day={01},
volume={71},
number={3},
pages={205-226},
issn={0044-3328},
doi={10.1007/BF01341708},
url={https://doi.org/10.1007/BF01341708}
}

@book{Nielsen00,
  author = {Nielsen, Michael A. and Chuang, Isaac L.},
  publisher = {Cambridge University Press},
  title = {Quantum Computation and Quantum Information},
  year = 2000,
  doi={https://doi.org/10.1017/CBO9780511976667}
}

@article{Creutz99,
  title = {End States, Ladder Compounds, and Domain-Wall Fermions},
  author = {Creutz, Michael},
  journal = {Phys. Rev. Lett.},
  volume = {83},
  issue = {13},
  pages = {2636--2639},
  numpages = {0},
  year = {1999},
  month = {Sep},
  publisher = {American Physical Society},
  doi = {10.1103/PhysRevLett.83.2636},
  url = {https://link.aps.org/doi/10.1103/PhysRevLett.83.2636}
}

@article{Baradaran18,
title = "Jastrow-like ground states for quantum many-body potentials with near-neighbors interactions",
journal = "Annals of Physics",
volume = "388",
pages = "147 - 161",
year = "2018",
issn = "0003-4916",
doi = "https://doi.org/10.1016/j.aop.2017.11.007",
url = "http://www.sciencedirect.com/science/article/pii/S0003491617303226",
author = "M. Baradaran and J. A. Carrasco and F. Finkel and A. Gonz\'alez-L\'opez"
}

@article{Yadav19,
title = "Rationally extended many-body truncated Calogero–Sutherland model",
journal = "Annals of Physics",
volume = "400",
pages = "189 - 197",
year = "2019",
issn = "0003-4916",
doi = "https://doi.org/10.1016/j.aop.2018.11.009",
url = "http://www.sciencedirect.com/science/article/pii/S000349161830294X",
author = "R. K. Yadav and A. Khare and N. Kumari and B. P. Mandal"
}

@article{Sutherland71,
author = {Sutherland, B.},
title = {Quantum Many‐Body Problem in One Dimension: Ground State},
journal = {Journal of Mathematical Physics},
volume = {12},
number = {2},
pages = {246-250},
year = {1971},
doi = {10.1063/1.1665584},
URL = {https://doi.org/10.1063/1.1665584}
}

@book{Sutherland04,
author = {Sutherland, Bill},
title = {Beautiful Models},
publisher = {World Scientific},
year = {2004},
doi = {10.1142/5552},
address = {},
edition   = {},
URL = {https://www.worldscientific.com/doi/abs/10.1142/5552}
}

@article{Olshanii98,
  title = {Atomic Scattering in the Presence of an External Confinement and a Gas of Impenetrable Bosons},
  author = {Olshanii, M.},
  journal = {Phys. Rev. Lett.},
  volume = {81},
  issue = {5},
  pages = {938--941},
  numpages = {0},
  year = {1998},
  month = {Aug},
  publisher = {American Physical Society},
  doi = {10.1103/PhysRevLett.81.938},
  url = {https://link.aps.org/doi/10.1103/PhysRevLett.81.938}
}

@article{Girardeau04,
title = "Effective interactions, Fermi–Bose duality, and ground states of ultracold atomic vapors in tight de Broglie waveguides",
journal = "Optics Communications",
volume = "243",
number = "1",
pages = "3 - 22",
year = "2004",
issn = "0030-4018",
doi = "https://doi.org/10.1016/j.optcom.2004.09.079",
url = "http://www.sciencedirect.com/science/article/pii/S0030401804010582",
author = "M.D. Girardeau and Hieu Nguyen and M. Olshanii"
}

@book{KBI97,
  title={Quantum Inverse Scattering Method and Correlation Functions},
  author={V. E. Korepin and N. M. Bogoliubov and A. G. Izergin},
  year={1997},
  publisher={Cambridge, Cambridge},
  doi={https://doi.org/10.1017/CBO9780511628832}
}

@book{Takahashi99,
  title={Thermodynamics of One-Dimensional Solvable Models},
  author={M. Takahashi},
  year={1999},
  publisher={Cambridge, Cambridge},
  doi={https://doi.org/10.1017/CBO9780511524332}
}

@article{McGuire64,
author = {McGuire,J. B. },
title = {Study of Exactly Soluble One‐Dimensional N‐Body Problems},
journal = {Journal of Mathematical Physics},
volume = {5},
number = {5},
pages = {622-636},
year = {1964},
doi = {10.1063/1.1704156},
URL = {https://doi.org/10.1063/1.1704156}
}

@article{Polychronakos93,
  title = {Lattice integrable systems of Haldane-Shastry type},
  author = {Polychronakos, Alexios P.},
  journal = {Phys. Rev. Lett.},
  volume = {70},
  issue = {15},
  pages = {2329--2331},
  numpages = {0},
  year = {1993},
  month = {Apr},
  publisher = {American Physical Society},
  doi = {10.1103/PhysRevLett.70.2329},
  url = {https://link.aps.org/doi/10.1103/PhysRevLett.70.2329}
}

@article{Polychronakos06,
doi = {10.1088/0305-4470/39/41/S07},
url = {https://dx.doi.org/10.1088/0305-4470/39/41/S07},
year = {2006},
month = {sep},
publisher = {},
volume = {39},
number = {41},
pages = {12793},
author = {Alexios P Polychronakos},
title = {The physics and mathematics of Calogero particles},
journal = {Journal of Physics A: Mathematical and General}
}

@article{Ares18,
  title = {Orthogonality catastrophe and fractional exclusion statistics},
  author = {Ares, Filiberto and Gupta, Kumar S. and de Queiroz, Amilcar R.},
  journal = {Phys. Rev. E},
  volume = {97},
  issue = {2},
  pages = {022133},
  numpages = {6},
  year = {2018},
  month = {Feb},
  publisher = {American Physical Society},
  doi = {10.1103/PhysRevE.97.022133},
  url = {https://link.aps.org/doi/10.1103/PhysRevE.97.022133}
}

@article{Fogarty20,
  title = {Orthogonality Catastrophe as a Consequence of the Quantum Speed Limit},
  author = {Fogarty, Thom\'as and Deffner, Sebastian and Busch, Thomas and Campbell, Steve},
  journal = {Phys. Rev. Lett.},
  volume = {124},
  issue = {11},
  pages = {110601},
  numpages = {7},
  year = {2020},
  month = {Mar},
  publisher = {American Physical Society},
  doi = {10.1103/PhysRevLett.124.110601},
  url = {https://link.aps.org/doi/10.1103/PhysRevLett.124.110601}
}

@article{Haldane88,
  title = {Exact Jastrow-Gutzwiller resonating-valence-bond ground state of the spin-$\frac{1}{2}$ antiferromagnetic Heisenberg chain with 1/${\mathrm{r}}^{2}$ exchange},
  author = {Haldane, F. D. M.},
  journal = {Phys. Rev. Lett.},
  volume = {60},
  issue = {7},
  pages = {635--638},
  numpages = {0},
  year = {1988},
  month = {Feb},
  publisher = {American Physical Society},
  doi = {10.1103/PhysRevLett.60.635},
  url = {https://link.aps.org/doi/10.1103/PhysRevLett.60.635}
}

@article{Kawakami1993,
  title = {Novel hierarchy of the SU(N) electron models and edge states of fractional quantum Hall effect},
  author = {Kawakami, Norio},
  journal = {Phys. Rev. Lett.},
  volume = {71},
  issue = {2},
  pages = {275--278},
  numpages = {0},
  year = {1993},
  month = {Jul},
  publisher = {American Physical Society},
  doi = {10.1103/PhysRevLett.71.275},
  url = {https://link.aps.org/doi/10.1103/PhysRevLett.71.275}
}

@book{Jain2007,
  author    = {Jainendra K. Jain},
  title     = {Composite Fermions},
  publisher = {Cambridge University Press},
  year      = {2007},
  address   = {Cambridge, UK},
  isbn      = {9780511607561},
  doi       = {10.1017/CBO9780511607561},
  url       = {https://www.cambridge.org/core/books/composite-fermions/AB22E09A3F9C4E98F91E1BB447AF5778}
}

@article{Shastry88,
  title = {Exact solution of an S=1/2 Heisenberg antiferromagnetic chain with long-ranged interactions},
  author = {Shastry, B. Sriram},
  journal = {Phys. Rev. Lett.},
  volume = {60},
  issue = {7},
  pages = {639--642},
  numpages = {0},
  year = {1988},
  month = {Feb},
  publisher = {American Physical Society},
  doi = {10.1103/PhysRevLett.60.639},
  url = {https://link.aps.org/doi/10.1103/PhysRevLett.60.639}
}

@article{Vojta2019,
   author = "Vojta, Thomas",
   title = "Disorder in Quantum Many-Body Systems", 
   journal= "Annual Review of Condensed Matter Physics",
   year = "2019",
   volume = "10",
   number = "Volume 10, 2019",
   pages = "233-252",
   doi = "https://doi.org/10.1146/annurev-conmatphys-031218-013433",
   url = "https://www.annualreviews.org/content/journals/10.1146/annurev-conmatphys-031218-013433",
   publisher = "Annual Reviews",
   issn = "1947-5462",
   type = "Journal Article",
  }

@article{Binder1986,
  title = {Spin glasses: Experimental facts, theoretical concepts, and open questions},
  author = {Binder, K. and Young, A. P.},
  journal = {Rev. Mod. Phys.},
  volume = {58},
  issue = {4},
  pages = {801--976},
  numpages = {0},
  year = {1986},
  month = {Oct},
  publisher = {American Physical Society},
  doi = {10.1103/RevModPhys.58.801},
  url = {https://link.aps.org/doi/10.1103/RevModPhys.58.801}
}

@article{delcampo16,
	doi = {10.1088/1367-2630/18/1/015014},
	url = {https://doi.org/10.1088%2F1367-2630%2F18%2F1%2F015014},
	year = 2016,
	month = {jan},
	publisher = {{IOP} Publishing},
	volume = {18},
	number = {1},
	pages = {015014},
	author = {Adolfo del Campo},
	title = {Exact quantum decay of an interacting many-particle system: the Calogero{\textendash}Sutherland model},
	journal = {New Journal of Physics}
}

@article{Sutherland98,
  title = {Exact Coherent States of a One-Dimensional Quantum Fluid in a Time-Dependent Trapping Potential},
  author = {Sutherland, Bill},
  journal = {Phys. Rev. Lett.},
  volume = {80},
  issue = {17},
  pages = {3678--3681},
  numpages = {0},
  year = {1998},
  month = {Apr},
  publisher = {American Physical Society},
  doi = {10.1103/PhysRevLett.80.3678},
  url = {https://link.aps.org/doi/10.1103/PhysRevLett.80.3678}
}

@article{delcampo08,
  title = {Fermionization and bosonization of expanding one-dimensional anyonic fluids},
  author = {del Campo, A.},
  journal = {Phys. Rev. A},
  volume = {78},
  issue = {4},
  pages = {045602},
  numpages = {4},
  year = {2008},
  month = {Oct},
  publisher = {American Physical Society},
  doi = {10.1103/PhysRevA.78.045602},
  url = {https://link.aps.org/doi/10.1103/PhysRevA.78.045602}
}

@article{Pittman17,
  title = {Truncated Calogero-Sutherland models},
  author = {Pittman, S. M. and Beau, M. and Olshanii, M. and del Campo, A.},
  journal = {Phys. Rev. B},
  volume = {95},
  issue = {20},
  pages = {205135},
  numpages = {9},
  year = {2017},
  month = {May},
  publisher = {American Physical Society},
  doi = {10.1103/PhysRevB.95.205135},
  url = {https://link.aps.org/doi/10.1103/PhysRevB.95.205135}
}

@article{JainKhare99,
title = "An exactly solvable many-body problem in one dimension and the short-range Dyson model",
journal = "Physics Letters A",
volume = "262",
number = "1",
pages = "35 - 39",
year = "1999",
issn = "0375-9601",
doi = "https://doi.org/10.1016/S0375-9601(99)00637-4",
url = "http://www.sciencedirect.com/science/article/pii/S0375960199006374",
author = "Sudhir R. Jain and Avinash Khare"
}

@article{Auberson01,
	doi = {10.1088/0305-4470/34/4/302},
	url = {https://doi.org/10.1088%2F0305-4470%2F34%2F4%2F302},
	year = 2001,
	month = {jan},
	publisher = {{IOP} Publishing},
	volume = {34},
	number = {4},
	pages = {695--724},
	author = {Guy Auberson and Sudhir R Jain and Avinash Khare},
	title = {A class of {N}-body problems with nearest- and next-to-nearest-neighbour interactions},
	journal = {Journal of Physics A: Mathematical and General}
}

@article{Basu-Mallick01,
title = "Exact solution of a many body problem with nearest and next-nearest neighbour interactions",
journal = "Physics Letters A",
volume = "279",
number = "1",
pages = "29 - 32",
year = "2001",
issn = "0375-9601",
doi = "https://doi.org/10.1016/S0375-9601(00)00816-1",
url = "http://www.sciencedirect.com/science/article/pii/S0375960100008161",
author = "B. Basu-Mallick and Anjan Kundu"
}

@article{Ezung05,
  title = {Quantum many-body systems with nearest and next-to-nearest neighbor long-range interactions},
  author = {Ezung, Meripeni and Gurappa, N. and Khare, Avinash and Panigrahi, Prasanta K.},
  journal = {Phys. Rev. B},
  volume = {71},
  issue = {12},
  pages = {125121},
  numpages = {8},
  year = {2005},
  month = {Mar},
  publisher = {American Physical Society},
  doi = {10.1103/PhysRevB.71.125121},
  url = {https://link.aps.org/doi/10.1103/PhysRevB.71.125121}
}

@article{Yang1967,
  title = {Some Exact Results for the Many-Body Problem in one Dimension with Repulsive Delta-Function Interaction},
  author = {Yang, C. N.},
  journal = {Phys. Rev. Lett.},
  volume = {19},
  issue = {23},
  pages = {1312--1315},
  numpages = {0},
  year = {1967},
  month = {Dec},
  publisher = {American Physical Society},
  doi = {10.1103/PhysRevLett.19.1312},
  url = {https://link.aps.org/doi/10.1103/PhysRevLett.19.1312}
}

@article{Yang2025,
  title = {Time-dependent Jastrow ansatz: Exact quantum dynamics, shortcuts to adiabaticity, and quantum quenches in strongly correlated many-body systems},
  author = {Yang, Jing and del Campo, Adolfo},
  journal = {Phys. Rev. A},
  volume = {111},
  issue = {5},
  pages = {053315},
  numpages = {23},
  year = {2025},
  month = {May},
  publisher = {American Physical Society},
  doi = {10.1103/PhysRevA.111.053315},
  url = {https://link.aps.org/doi/10.1103/PhysRevA.111.053315}
}

@article{Li:1995wr,
  title = {\ensuremath{\delta}-function spin-1/2 fermions in a one-dimensional potential well},
  author = {Li, You-Quan and Dai, Jian-Hui},
  journal = {Phys. Rev. A},
  volume = {53},
  issue = {6},
  pages = {3743--3748},
  numpages = {0},
  year = {1996},
  month = {Jun},
  publisher = {American Physical Society},
  doi = {10.1103/PhysRevA.53.3743},
  url = {https://link.aps.org/doi/10.1103/PhysRevA.53.3743}
}

@misc{Theorem_mondal2025,
      title={When Are Standard Graph Products Isomorphic?}, 
      author={Priti Prasanna Mondal and M. Rajesh Kannan and Fouzul Atik},
      year={2025},
      eprint={2508.04137},
      archivePrefix={arXiv},
      primaryClass={math.CO},
      url={https://arxiv.org/abs/2508.04137}, 
}

@book{Ushveridze1994,
  author    = {A. G. Ushveridze},
  title     = {Quasi-Exactly Solvable Models in Quantum Mechanics},
  edition   = {1st},
  publisher = {CRC Press},
  address   = {Boca Raton, FL, USA},
  year      = {1994},
  doi       = {10.1201/9780203741450},
  isbn      = {978-0203741450}
}

\end{document}